\documentclass[prl,superscriptaddress,reprint,amsmath,amssymb,graphicx,longbibliography]{revtex4-1}

\usepackage{graphicx}
\usepackage{sidecap}
\usepackage{xcolor}
\usepackage{makecell}
\usepackage{float}
\usepackage{hyperref}
\usepackage{bm}
\usepackage[utf8]{inputenc}
\usepackage{multirow,array}

\begin{document}

\title{Inertial Spinner Swarm Experiments: \\ Spin Pumping, Entropy Oscillations and Spin Frustration}
\author{Shengkai Li}
\thanks{These authors contributed equally to this work.}
\affiliation{%
Department of Physics, Princeton University, Princeton, New Jersey 08544, USA}%
\affiliation{%
Center for the Physics of Biological Function, Princeton University, Princeton, New Jersey 08544, USA}%

\author{Trung V. Phan}
\thanks{These authors contributed equally to this work.}
\affiliation{%
Yale University, New Haven, Connecticut 06520, USA}%

\author{Gao Wang}
\affiliation{%
Wenzhou Institute, University of Chinese Academy of Sciences, Wenzhou, Zhejiang 325000, China}%
\affiliation{%
School of Physical Sciences, University of Chinese Academy of Sciences, Beijing 100049, China}%

\author{Ramzi R. Khuri}
\affiliation{%
Department of Natural Sciences, Baruch College, City University of New York, New York, New York 10010, USA}%

\author{Robert H. Austin}
\thanks{austin@princeton.edu}
\affiliation{%
Department of Physics, Princeton University, Princeton, New Jersey 08544, USA}%

\author{Liyu Liu}
\thanks{lyliu@cqu.edu.cn}
\affiliation{%
Chongqing Key Laboratory of Soft Condensed Matter Physics and Smart Materials,College of Physics, Chongqing University, Chongqing 401331, China}%

\date{\today{}}

\begin{abstract}
  We present here an inertial active spinning swarm consisting of mixtures of opposite handedness torque driven spinners floating on an air bed with low damping. Depending on the relative spin sign, spinners can act as their own anti-particles and annihilate their spins.  Rotational energy can become highly focused, with minority fraction spinners pumped to very high levels of spin angular momentum. Spinner handedness also matters at high spinner densities but not low densities: oscillations in the mixing spatial entropy of spinners over time emerge if there is a net spin imbalance from collective rotations. Geometrically confined spinners can lock themselves into frustrated spin states. 

\end{abstract}

\maketitle

Active matter has its roots in biology, but active matter also occurs in physics. Challenges arise when the active matter is both inertial and chiral. Our inertial spinner active swarm is a form of active inertial matter \cite{iam1, iam2, iam3, te2022microscopic,li2022robotic}, where constituents with driven dual spin degrees of freedom exchange both translational kinetic energy and generalized angular momentum. While there have been pioneering simulation studies of active spinners at very high densities \cite{scooped}, and pioneering experiments in bacterial collective dynamics have seen some of the phenomena we observe here \cite{goldstein, libchaber} we believe these experiments are original in combining multiple handedness driven spinners as a function of both density and fractional handedness fractions, with surprising results.

We achieved low inertial damping in the equations of motion by floating actively rotational driven disks (spinners) on an air table.  Energy is fed into the system in two ways: (1) incoherent translational drive via the turbulent air flow of the air table; (2) constant torque spin drive on each disc by opposing battery powered air blowers on each disc. The generalized spin of our spinners is boolean in sign but analog in magnitude. Rotational degrees of freedom are coupled to translational degrees of freedom in collisions via teeth on the perimeters of the disks. Teeth-teeth interactions provide for strong translational coupling in collisions between spinners in a rather non-intuitive manner depending on the relative signs of the spin vectors of the colliding spinners. For ease of notation in what follows, we will call a counterclockwise (CCW) spinner as having $+$ spin, and a clockwise (CW) spinner has having $-$ spin. This separation into two distinct species is an unique feature of low-dimensional topology, since these species are equivalent up to a rotational transformation at higher dimensions.

\begin{figure}[!htb]
\centering
\includegraphics[width=0.5\textwidth]{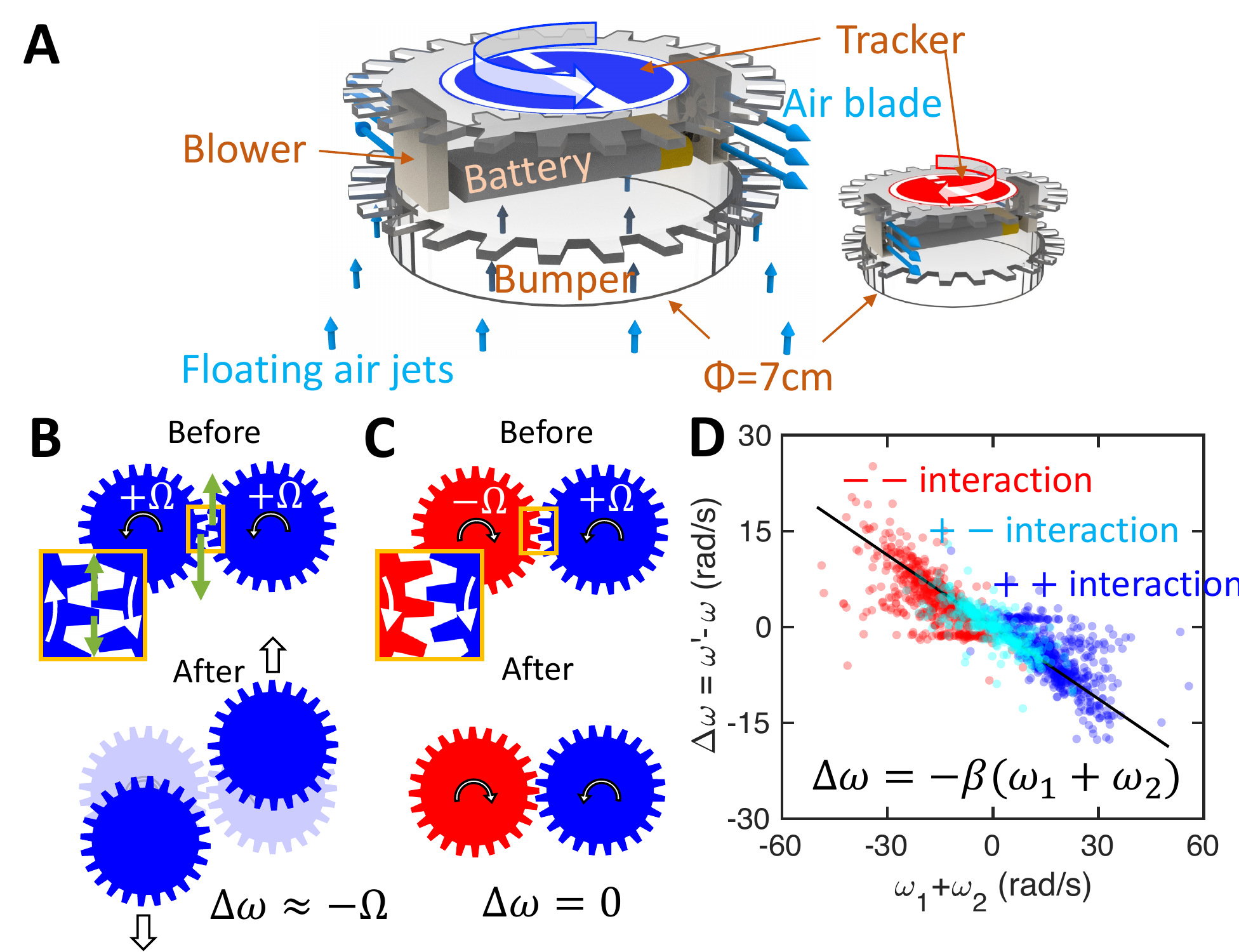} 
\caption{\textbf{Basics of the spinners.} (A) A schematic of a spinner floating on an air table. Air is ejected from two blowers on the sides of the geared discs in opposite direction thus rotating the spinner. A tracker with a white line and a binary barcode is mounted at the top of the spinner to track its rotational and translational motion. Blue and red is used for CCW and CW spinners respectively. A bumper (a petri dish) at the bottom elevates the spinner to avoid collision between the boundary wire and the blowers. (B) Two spinners with same spins repel each other due to opposite motion of gear teeth. From the collision, rotational energy is converted into translational energy. (C) Two spinners with opposite spins have teeth motion in the same direction. See \href{https://www.dropbox.com/s/qbyxp9zg0iblhmq/SI1_ffmpeg.mp4?dl=0}{SI1.mp4} for sample collisions. (D) Change of angular velocity of both spinners is proportional to the sum of both velocities before collision (Eq.\ref{alphabet_of_collision}). Here $\Delta\omega=\omega'-\omega$ uses the average of both spinners $\Delta\omega=(\Delta\omega_1+\Delta\omega_2)/2$. The black theory line uses the inertial property $\beta\sim 3/8$ measured from experiments (see SM section II A \cite{schedlinski2001survey}). The color dots show the result from 10$^4$ collision events in experiments.}
\label{cartoon}
\end{figure}

 Colliding spinners with the same sign of spin vector have tangential velocity vectors which are opposed in sign and transfer spin angular momentum into orbital angular momentum, and annihilate their spins. However collisions between spinners of opposite sign tend to maintain their spin vectors since the tangential velocity vectors of the teeth are in the same direction, and transfer little rotational energy into orbital angular momentum \cite{scholz2018rotating,banerjee2017odd}. See Fig. \ref{cartoon} for a pictorial description.

We derive in the Supplementary Material the spin interchange averaged all impact parameters $b$ for a two-body spinner collision  of spinners of radius $R$, mass $M$ and moment of inertia $I$ with initial angular velocities $\omega_1$ and $\omega_2$ respectively and exiting angular velocities  $\omega'_1$ and $\omega'_2$:
\begin{equation}
\omega_1' - \omega_1 = \omega_2' - \omega_2 = -\beta \left( \omega_1 + \omega_2 \right) \ .
\label{alphabet_of_collision}
\end{equation}
where $\beta=1/2~(1+I/MR^2)<1/2$ (see Fig.\ref{cartoon}D for experiment data).

Eq. \ref{alphabet_of_collision} has unexpected predictions which we exploit in the following 2 experiments.\\

{\bf Spin Pumping of Minority Handedness:}. If both spinners have the same sign for the initial spin (a $++$ collision) they {\em lose} spin angular momentum which gets converted to orbital (translational) angular momentum, while spinners of opposite sign (a $+-$ collision) {\em maintain} spin angular momentum depending on the relative magnitudes on the angular velocities, with zero loss if $\omega_1 = - \omega_2$! Since orbital angular momentum is a form of translational kinetic energy, one would expect that an equal mixture of spin up and spin down spinners would have a high spin effective temperature but a low translational temperature, while a population of all spin up or spin down spinners might have a low spin temperature but a high translational temperature.

\begin{figure*}[!thb]
\centering
\includegraphics[width=0.8\textwidth]{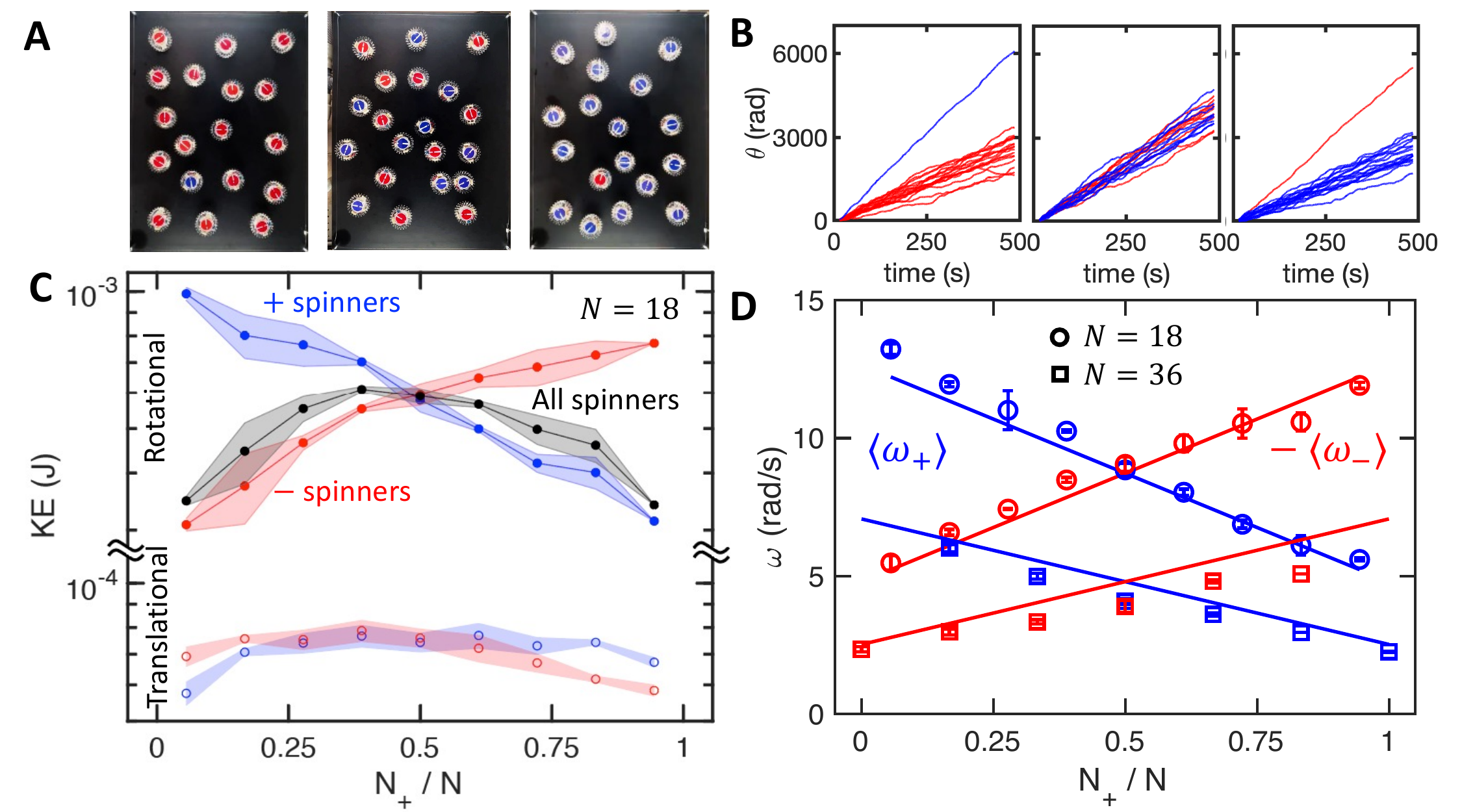} 
\caption{\textbf{Effects from the spin-up/spin-down population ratio.} (A) Three typical mixture of the up and down spinners. (B) The evolution of self spin in spin-down dominant, even mixed, and spin-up dominant cases. The blue and red curves shows the self spin temporal evolution of all $18$ spinners. See \href{https://www.dropbox.com/s/upmd9s47xk7fssc/SI2_ffmpeg.mp4?dl=0}{SI2.mp4} for sample movies. (C) Translational (open dots) and rotational energies (solid dots) per spinner averaged over time for various spin mixture ratios over $18$ spinners. Blue and red curves show the energies of the spin-up (CCW, blue) and the spin-down (CW, red) respectively while the green curve shows the overall statistics. (D) Measured spin rate (error bars) compared with theory (Eq. \ref{spin_pump_eq}, solid lines) using parameters $\Omega = 32$ rad/s and $\alpha = 6$ measured from individual spinners (see SM section III B).}
\label{figFraction}
\end{figure*}

This phenomena of dramatic minority spin-pumping is experimentally seen in Fig. \ref{figFraction}B. Experiments were carried out as a function of the density of the spinners on the air table. At low spinner density most collisions are binary in nature and Eq. \ref{alphabet_of_collision} can be used to predict transfer of spin angular momentum which is constantly being pumped in by the tangentially configured blowers into translational kinetic energy, which is either partially lost in the inelastic collisions of the spinners and what remains is eventually dissipated via viscous drag of movement of the spinners.

We applied both theory and simulation to understand these phenomena. Conservation of angular momentum gives the fundamental physics of collision as shown by Eq. \eqref{alphabet_of_collision}, which is in good agreement with data of collisions extracted from experiments (see Fig. \ref{figFraction}D). Based on this, we can create an inertial-dominated toy-model which allows us to write down the equation of rotational motion for each spinners, as we show in SM Section III \cite{kardar2007statistical,taloni2015collisional,taylor1717methodus,thomas1968calculus,phan2021doesn}. After time-averaging many collisions for all spinners in each species, we arrive at the estimation for the average spinning velocity $\langle \omega_{\pm} \rangle$ from the population number $N_\pm$:
\begin{equation}
\langle \omega_\pm \rangle = \pm \Omega \ \times \\
\frac{\alpha \left( \alpha + 3N - 2 - 2N_\pm \right)}{(\alpha + 2N-2)(\alpha + N -2)}\ ,
\label{spin_pump_eq}
\end{equation}
where $\alpha$ depends on the inertial properties of spinners and the driving torques generated by the air-blowers, $\Omega$ is the maximum angular-velocity the spinners can possibly be. For a fixed value of $N$ then $\langle \omega_\pm \rangle$ monotonically decreases as $N_\pm$ increases: the minority will always spin faster than the majority! As shown in Fig. \ref{figFraction}C, this model matches with experimental observations. Simulations using parameters measured from experiments also showed agreement with the results on energy and mixing dynamics' dependence on the spin ratio (Fig.S15). Our simulation shows the concave geometry alone can generate the tangential interaction between spinners without using friction as a substitute in simulation \cite{liu2020oscillating}. As expected, after we remove all dissipation forces (translational and rotational aerodynamic drag) and energy injection (rotational torque from the air blowers on the spinners), the translational and rotational energy show equipartition of energy in simulation (see SM section VII B and \href{https://www.dropbox.com/s/w4yb21ikjhta5c8/SI5_ffmpeg.mp4?dl=0}{SI5.mp4}).

The difference between collisions of the same- and opposite-handedness spinner pairs creates different emergent spin rate distributions and spatial currents depending on the ratio between the left-handed and right-handed spinners. The concave-down behavior of the total rotational energy centered at $n_+=1/2$ and the asymmetry of the translational energies as functions of species populations (see Fig. \ref{figFraction}) can be captured by crude estimations made in SM section IV.

Fig. \ref{figFraction} shows the rotational and translational energy/spinner for $N=N_+ + N_-$ = 18 spinners as a function of the $n_+ = N_+$/$N$ in steady state. There are several striking aspects to this data. (1) Clearly at extrema/minima values of $n_+$ localization of the spin energy in the minority fraction is very clear; (2) At extrema/minima values of $n_+$ the high rotational energy of the minority spin substantially deducts rotational energy from the overall per spinner average rotational energy; (3) The average translational energy of the system per spinner is close to 1/10 the average rotational energy of the spinners, agreeing with the finding by Nguyen et al \cite{nguyen2014emergent} at $\phi=0.16$, the density of the experiment in Fig.\ref{figFraction}; (4) At extrema/minima values of $n_+$ the spin-pumped spinners also extract translational energy from the spinners.\\

{\bf Spatial Entropy Oscillation in Mixing} An interesting aspect of this form of inertial active matter is the dependence of the spatial flow of mixing (positional) entropy on the net handedness of the mixture of spinners.  Since there is a flow of energy from the highly localized spins of the spinners to translational energy and there is a strong dependence of this flow on the relative spins of the colliding spinners, one would expect that there would be strong dependence of entropic mixing times on the net handedness of the spinner mixture, but we show an additional unexpected collective rotation of the spinners which gives rise to entropy oscillations.

\begin{figure*}[!thb]
\centering
\includegraphics[width=0.74\textwidth]{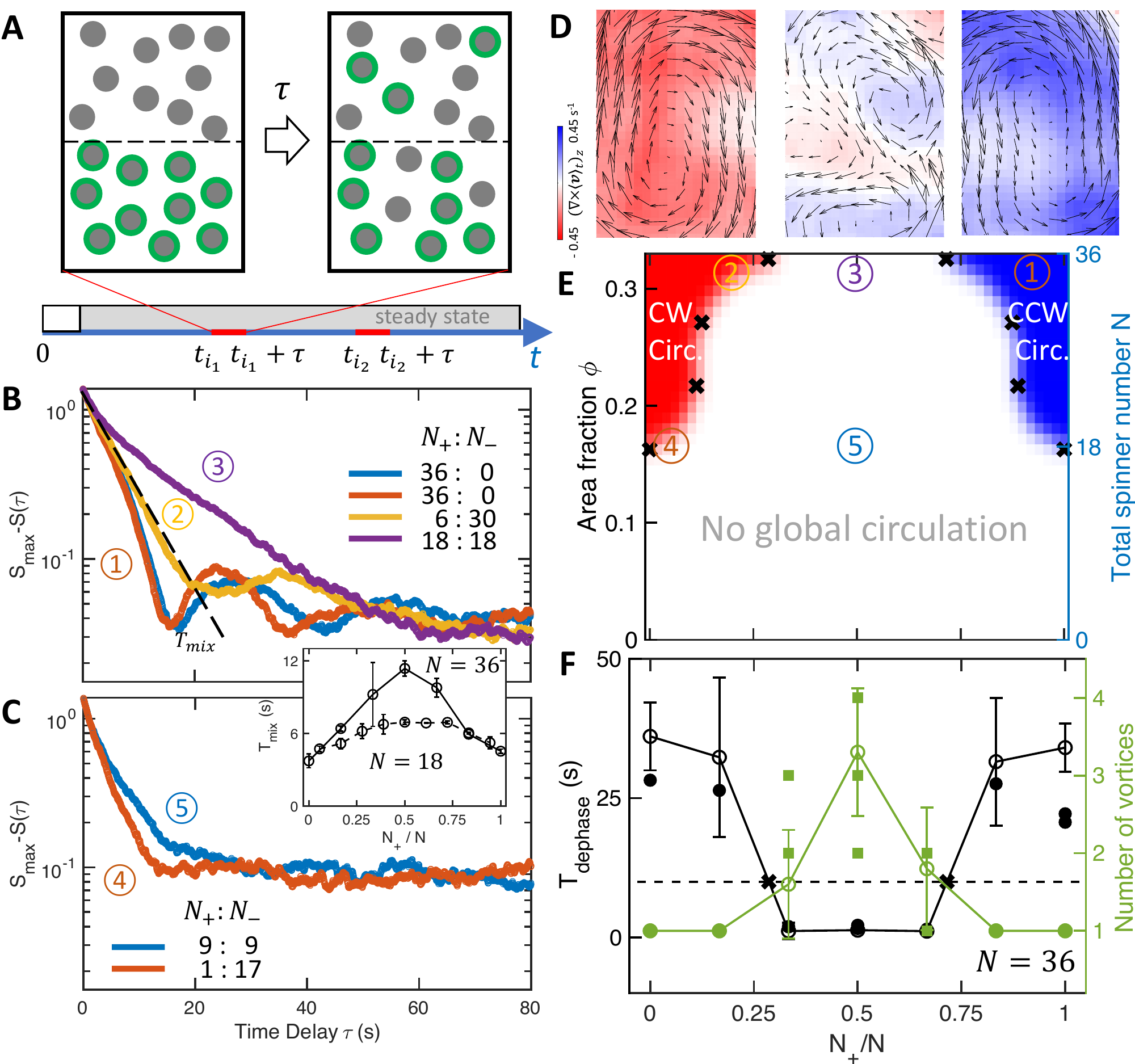} 
\caption{\textbf{Entropy production.} (A) To evaluate the entropy change in time $\tau$, we mark spinners on both sides (lower part in green circles as an example here) at the beginning of a time interval and evaluate the entropy after $\tau$. One can find many such evolutions with time interval $\tau$. We average the ensemble of these intervals for the entropy $S(\tau)$. See movie \href{https://www.dropbox.com/s/t96o32yb8btcos6/SI3_ffmpeg.mp4?dl=0}{SI3.mp4} for visualization. (B) Entropy approaches maximum entropy over time for $36$ spinners. The decay time for the incipient drop is used as the characteristic time for mixing ($T_{mix}$). (C) Entropy over time for $18$ spinners. Inset: Mixing time dependence on the spin ratio for both densities. (D) Time-averaged velocity of spinners for pure-bottom-spin, even, and pure-upper-spin collectives for 36-spinner collectives over 500 seconds. (E) Three distinct phases of the spinner collectives depending on the spin ratio and area fraction. The boundary of phases are determined by the sharp transition of dephasing time from simulation. (F) Dephasing time ($T_{dephase}$, the characteristic time for the oscillation part of entropy increase to decay) and number of emergent vortices at different spin ratios for 36-spinner collectives. The error bars and solid dots show the simulation and experiment respectively. The phase boundary in (E) uses $T_{dephase}=10$ s as shown by crosses.}
\label{figEntropy}
\end{figure*}
\twocolumngrid

 Note that all our spinners are identical (other than the sign of their torque drive) but distinguishable due to the code written on each one! Mixing entropy was computed by tracking individual spinners as to their position in the upper $u$ and bottom $b$ position over time, where we divided our spinners by their position in the upper $u$ and bottom $b$ region of the table starting at $t=t_0$.The positional entropy $S$ of an ensemble of spinners at a given time is then given by:
\begin{equation}
S(t_0\rightarrow t) = - \left [ p^u_u \ln p^u_u + p^b_u \ln p^b_u + p^b_b \ln p^b_b + p^u_b \ln p^u_b      \right ]
\end{equation}
where $p^u_u$ is the joint probability of finding a spinner originally in the up location still in the up location at time $t$, $p^b_u$ is the joint cross probability of finding a spinner originally in upper location now in the bottom location, etc.  At $t= \infty$ for a uniform mixture $S(t_0 \rightarrow\infty)= S_{max} = 4 \times -[ 0.5 \ln (0.5)] = 2 \ln 2$, while at $t=t_0, S(t_0\rightarrow t_0) = 0$ since all the cross joint probabilities are $0$. Entropy $S(\tau)$ was measured by marking the spinners in the two sides of the arena (Fig.\ref{figEntropy}A, \href{https://www.dropbox.com/s/t96o32yb8btcos6/SI3_ffmpeg.mp4?dl=0}{SI3.mp4}) and sampling over increasingly separated in time placements of the spinners:
\begin{equation}
S(\tau)=\langle S(t_i\rightarrow t_i+\tau)\rangle_i.
\end{equation}

We evaluated the difference between the entropy $S(\tau)$ and the maximum $S_{max}$ at steady state after cutting the initial transient part when the air table gas jets are activated.

$S(\tau)$ generally increases with time as is expected for mixing. However, when the spinner density is high enough, we see dephasing oscillations in $S(\tau)$ with time as well for spinner mixtures which have a net initial global spin. This is due to transfer of the net global spin to a global net orbital angular momentum, so that a circulation of the spinners transiently exists due to the inertial nature of of the motion. This topological edge current at the outer boundary (see Fig.\ref{figEntropy}D for the current) is also observed in other systems with rotating objects \cite{yang2020robust,petroff2023density}. These oscillations do not appear for balanced initial spin states, or at low spinner densities (Fig.\ref{figEntropy}E). For the 36-spinner experiments (area fraction $\phi=0.32$), the spin ratio interval for no-oscillation is $\mid n_+ - 1/2\mid < \Delta \approx 1/4$. For the 18-spinner experiments (area fraction $\phi=0.16$), $\Delta$ is almost $1/2$. We show how the oscillation dephasing of the mixing entropy can emerge in SM section V.

\begin{figure*}[!thb]
\centering
\includegraphics[width=0.8\textwidth]{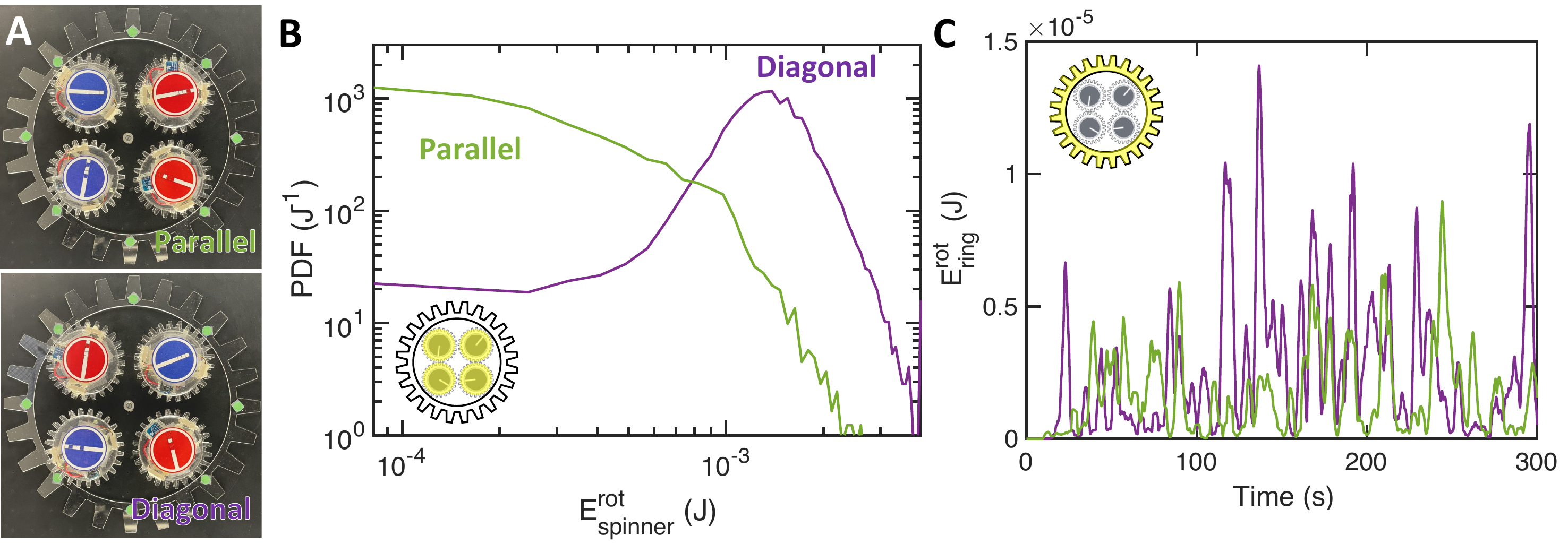} 
\caption{\textbf{Confined spinners.} (A) Two topologies: parallel and diagonal (Movie \href{https://www.dropbox.com/s/i3opuwfvxk2j154/SI4_ffmpeg.mp4?dl=0}{SI4.mp4}).
(B) Probability distribution of rotational energies for the two topological edge configurations. (C) Transfer of spin energy to the outer floating ring for the two topologies.}
\label{topology}
\end{figure*}

Another surprising result is that the characteristic mixing time $T_{mix}$ for the entropy to relax to the maximum value peaks at even spin mixtures and is lowest when the spinners only have one species (Fig.\ref{figEntropy}C inset). This implies that initial even mixed spin states jam more easily presumably because little spin energy is transferred into translational orbital energy for like-spin collisions. Further, as the spin ratio $n_+$ approaches the region where the increase of mixing entropy does not oscillate, the number of vortices increases critically from $1$ (one unique global circulation) while the overall vorticity remains neutral since the positive vortices pair with the negative vortices. The positions of the vortices move over time and the vary over different experiments. These features are also observed in simulations (Fig.S5). Once the spin ratio $n_+$ is within the critical ratio, the vortices are much more local and motile, presumably slowing down the mixing process. We posit a two-species generalization of the field theory of spinners \cite{tsai2005chiral,yang2020robust} would find the criticality of spin ratio and the emergence of vortices as approaching to the boundary of different regimes. The entropy oscillation reveals that the system time-dependent mixing kinetics is like a ``healing'' process for the entire system, which optimizes the spatial configurations of the spinners inside.\\

{\bf Spin Frustration:} The influence of topology and spinner placement is informative.  In a simple example of how topology and edge state placement greatly changes the spinner dynamics, we placed 4 spinners within a floating circle of inner diameter $D$ equal to 6 spinner radii, such that while the spinners could rotate under the applied blower torque and translate enough to freely collide, they could not exchange center of mass positions (see Fig. \ref{topology}).  Under those conditions there are only two possible topologies for a zero net spin collective: 2 spinners side by side of like spin, or diagonally opposed.

The spinner dynamics became quite different because in the side-by-side topology $++$ spin annihilating collisions are allowed and thus rotational kinetic energy is often lost to translational kinetic energy, while in the diagonally opposed configuration only $+-$ spin collisions are allowed, and thus high spin angular momenta should occur since loss of spin energy is minimized. Fig. \ref{topology} shows this to be the case: the parallel topology on average has low spin kinetic energy with a power-law like probability distribution function, while the diagonal topology has a steady state of high spin kinetic energy with a peaked probability distribution function. 

The floating confining ring can act as a transmission connecting spin dynamics within the ring to external rings, so that in principle our spinner can become a scale-free fractal form of active matter by connecting to every increasing sized toothed rings. As a first step demonstration of this, in Fig. \ref{topology}C we observe that the floating ring dynamics are strongly determined by the topology of the spinners contained within the ring. The intermittency of observed rotational energies indicate that this geometrically confined system of spinners is in frustration \cite{ramirez2003geometric,ramirez2001geometrical,spin-ice}, as there exists no steady state. We study other possible arrangements of 4 spinners in SM section VI, and also give an example for a scale-free fractal gear design there. Future study would include designs of fractal gears in more levels, which could bring more complex and interesting spatial-temporal dynamics.



Our findings show how energy and entropy flow through the translational and rotational degrees of freedom via inertial interactions in an ensemble of spin-up and spin-down spinners. We discovered a spin pumping mechanism which focuses rotational energy on the spinner species with smaller population numbers, and the optimum spin-up/spin-down ratio for highest translational energy of each spinner species. We observed the recurrence of mixing entropy at high density and purity of spinners, which explain with a topological edge state \cite{hasan2010colloquium} where fast circulating outer-flow is accompanied by slow mixing inner-core of spinners. We considered the inertial interactions between spinners which can be viewed as the memory chains between collisions, and have solved this system for the spinning motion at least in the mean-field approximation.

The inertial active matter introduced here exhibits many exotic emergent phenomena beyond conventional statistical mechanics, such as the expected violation of equipartion \cite{equipartition} and more surprisingly transiently the second law of thermodynamics. If we enable field-mediated interactions \cite{phan2021bootstrapped,li2022field} by replacing air with a more viscous liquid medium, spinners could become two-dimensional vortex-sources and form stable rotating crystal structures \cite{aref2003vortex,nguyen2014emergent,li2021programming} which poses many unsolved puzzles in number theory and classical mathematics \cite{aref1998point,aref2007point}. In regards to topological restrictions, the inertial active matter could provide a ``gearbox'' foundation for a nested architecture of scale-free fractal machines \cite{kriegman2021scale}, which can be operated and controlled similar to how larger scale emergent robots can be created out of robot swarms \cite{savoie2019robot,boudet2021collections,li2021programming} with many possible  behaviors driven by the complex dynamics of the topology they are moving on \cite{wang2021emergent,wang2022robots}. Although our inertial active matter exhibits complex and counter-intuitive behavior, this technology is not complex and can be easily implemented even at the middle school level with a 3D printer, a smart phone, and an air hockey table. 

\section{Acknowledgement}

This work was supported by the The National Natural Science Foundation of China (11974066 and 12174041) and the US National Science Foundation (PHY-1659940 and PHY-1734030). We acknowledge useful discussions with Truong H. Cai, Huy D. Tran, Khang V. Ngo, Neymar da Silva, Junang Li, Endao Han, and the xPhO Discord group.

\bibliography{spinner}
\bibliographystyle{apsrev4-2}

\renewcommand{\thefigure}{S\arabic{figure}}
\renewcommand{\thetable}{S\arabic{table}}
\setcounter{figure}{0}
\onecolumngrid
\section*{Supplementary Material}
\subsection{I. Experiment setup}

The spinners consist of toothed acrylic discs which are laser-cut to have 24 teeth. The spinners were driven by 2 oppositely directed blowers (SUNON UB3-500B), driven at 3.7 volts by a LiIon 400 mA-hr battery (Adafruit 3898).  The battery was sandwiched between 2 of the toothed discs, so that change in handedness of the spin could be accomplished by simple inversion of the disk. To ensure the consistency of the rotational driving torque, we made sure the batteries did not run more than 30 minutes after being charged to full.

A 60mm-diameter plastic Petri dish (Falcon Plastics) was used to lift the tooth wheel so that the close to elastic collisions with the taut wire of the air-table did not involve the teeth and hence little spin angular momentum change. An iPhone camera running at 30 frames/sec was used to take continuous movies of the dynamics of the spinner active matter. A bar-code imprinted on the top of each spinner allowed us to keep track of the center of mass positions of individual spinner versus time, measured rotation of the bar-code about the center of mass allowed us to know the spin angular momentum of each spinner as a function of time.

Single spinners where studied to determine the effective rate of flow of translational energy into a spinner due to air turbulence, the effective translational damping coefficient $\eta$, the applied torque $\tau_{drive} = 3.4\times 10^{-5}$ N~m of the mounted blowers and the the rotational damping coefficient $\eta_{\varphi}= 1.01\times 10^{-6}$ N~m~s.  The mass of each spinner was $0.025$ kg, and the moment of inertia about the center of mass $I_{cm}$ was determined by measuring the oscillation frequency $\omega$ for small angular displacements a known distance $r$ from the center of mass (See Sec. II A for details). We determined $I_{cm}$ to be $1.02 \times 10^{-5}$ kg~m$^2$.

\subsection{II. Determination of Physical Parameters}
\subsubsection{A. Determination of the Spinner Moment of Inertia}\label{sec:inertia}

\begin{figure}[!htb]
\centering
\includegraphics[width=0.4\textwidth]{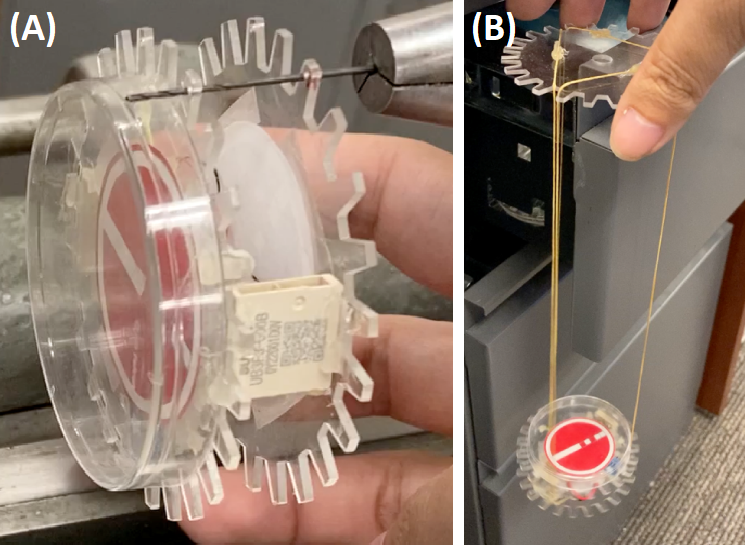}
\caption{(A) Pivot axis pendulum method. (B) Trifilar pendulum method.}
\label{figS00}
\end{figure}

We let the spinner of mass $M$ oscillates around a fixed pivot axis in the horizontal plane (see Fig. \ref{figS00}A). The axis is located at distance equal to the inner radius (gear teeth excluded) $R_{in}$ away from the center of the spinner. The moment of inertia of this physical pendulum $I_{piv}$ is given by:
\begin{equation}
I_{piv} = \kappa_{piv} MR_{in}^2 \ , \ \kappa_{piv} = \left( \frac{T^{(1)}_{piv}}{2\pi} \right)^2 \frac{g}{R_{in}} - 1 \ ,
\end{equation}
where $g$ is the gravitational acceleration and $T^{(1)}_{piv}$ is the time-period of the oscillation. 

Direct measurements give $M=0.025$kg and $R_{in}=0.03$m. Using $g=9.8$m/s$^2$, and from 5 measurements of 20-cycle time $T^{(20)}_{piv} = 20 T^{(1)}_{piv} = 8.6 \pm 0.2$s, we obtain $\kappa_{piv} = 0.52 \pm 0.07$. Note that the spinner radius (gear teeth included) is $R=0.035$m, therefore:
\begin{equation}
I_{piv} = MR^2 \ \times \ \left(  0.38 \pm 0.06 \right) \ \ \sim \ \ MR^2 \ \times \ 1/3 \ .
\label{pivot_finding}
\end{equation}
Measurement using the trifilar pendulum method \cite{schedlinski2001survey} also yields a similar result (see Fig. \ref{figS00}B).

\subsubsection{B. Air-Flow and Drag}

\begin{figure}[!htb]
\centering
\includegraphics[width=0.6\textwidth]{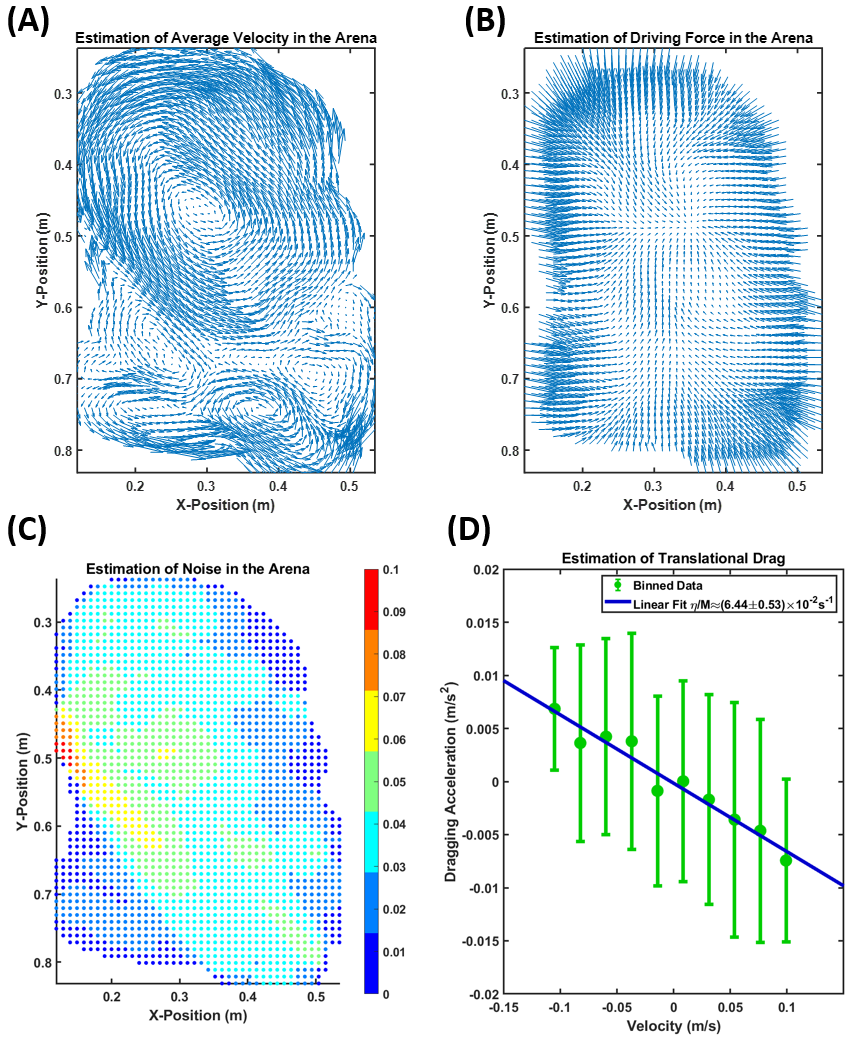}
\caption{(A) The vector-profile of the average spinner velocity $\langle \vec{V} \rangle$. (B) The vector-profile of the average air driving-force $\vec{f}/M$. (C) The average noise strength $\sqrt{\langle \vec{\xi}^2 \rangle/M^2}$. (A) Estimation of the drag $\eta_{trans}/M$ by looking at the relationship between the average spinner velocity and the dragging acceleration (which is just the acceleration but exclude the contribution from the driving-force).}
\label{figS07}
\end{figure}

It should be noted the air table has a very non-trivial flow-profile, therefore it is worthwhile to obtain some statistics about it. We assume that the interaction between the air bed and a spinner at position $\vec{X}(t)$ results in a total force $\vec{F}$ which has a spatial-dependence average $\vec{f}(\vec{x})$, random fluctuation $\vec{\xi}(x,t)$ and a drag $-\eta_{trans} \vec{V}$:

\begin{eqnarray}
M \frac{d^2}{dt^2} \vec{X}(t) = \vec{F}(t) = \vec{f}(\vec{x})\Big|_{\vec{x} = \vec{X}(t)} + \vec{\xi}(x,t) -\eta_{trans} \vec{V}(t) \ , \\
\vec{V}(t) = \frac{d}{dt} \vec{X}(t) \ . 
\end{eqnarray}
From the trajectories of single spinners in the arena, we can estimate $\langle \vec{V} \rangle$, $\vec{f}/M$, $\sqrt{\langle \vec{\xi}^2 \rangle/M^2}$ and $\eta_{trans}/M$ for any given position $\vec{x}$. The results does not seem to depend on whether the spinners are spin-up, spin-down or passive (blowers off), and are given in Fig. \ref{figS07}.

\subsection{III. Spinner Interaction and its Collective Effect}
\subsubsection{A. Pairwise Interaction of Spinners}
\begin{figure}[!htb]
\centering
\includegraphics[width=0.5\textwidth]{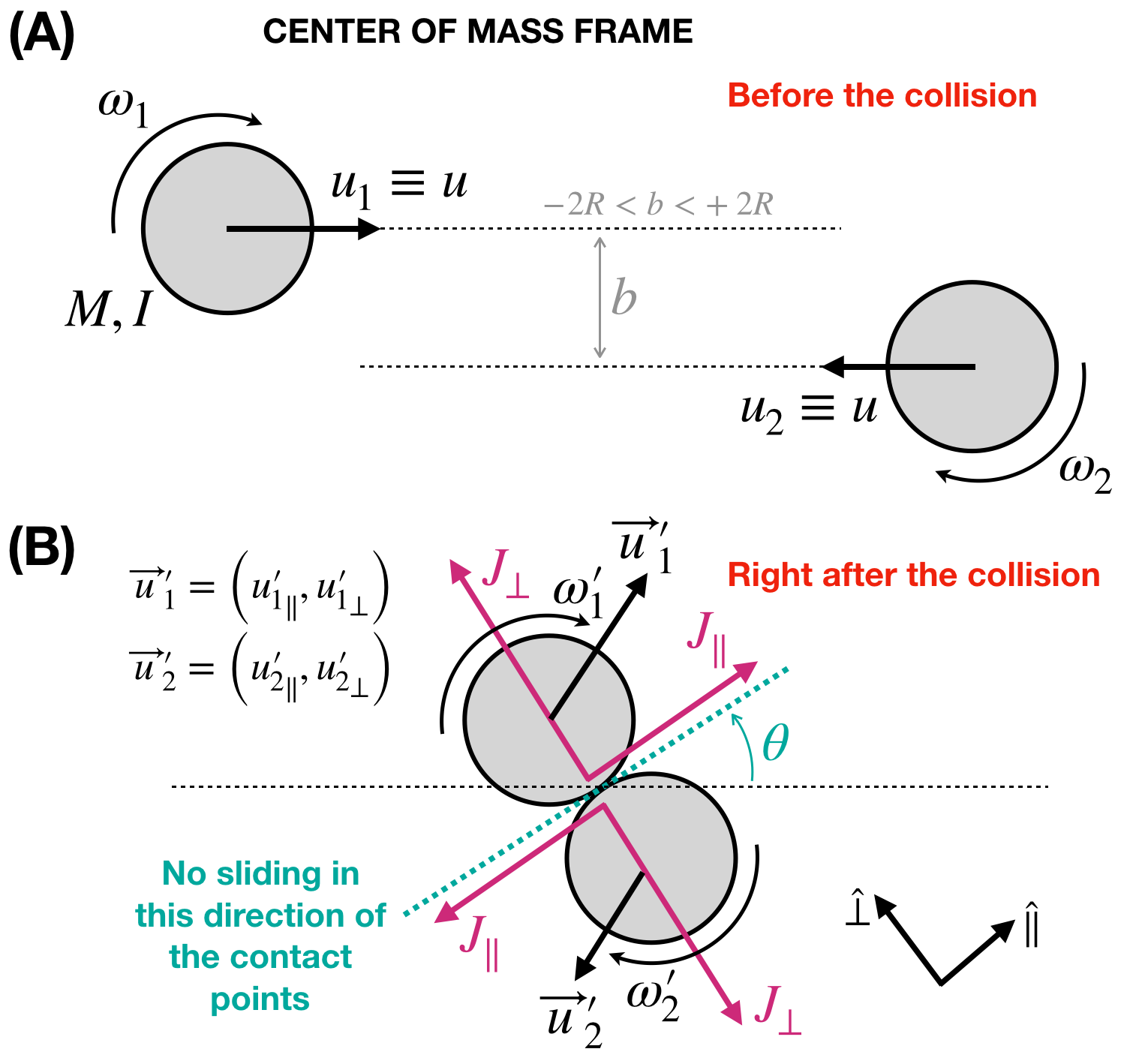}
\caption{In the center-of-mass frame of two spinners: (A) Before the collision. (B) Immediately after the collision.}
\label{figS01}
\end{figure}

We need to understand the reciprocal interactions between the spinners. Consider a collision between two spinners (of radius $R$, mass $M$, and moment of inertial $I$) having angular velocities $\omega_1$ and $\omega_2$ in their center-of-mass frame so that they travel at the same velocity $u$ but in opposite directions, with impact parameter $b$ (see Fig. \ref{figS01}A). After the collision, the total impulse they each receive is $J_\parallel$ and $J_\perp$ in the parallel direction $\hat{\parallel}$ and the perpendicular direction $\hat{\perp}$ with respect to the tangent of their contact points, so that their velocities vector and angular velocities become $\vec{v}_1' = \left({v_1'}_{\parallel},{v_1'}_{\perp}\right)$, $\vec{v}'_2=\left({v_2'}_{\parallel},{v_2'}_{\perp}\right)$ and $\omega_1'$, $\omega_2'$ (see Fig. \ref{figS01}B). Thus, define $\theta = \arcsin(b/2R)$, we have:
\begin{equation}
{u_1}_\parallel = -{u_2}_\parallel = u\cos\theta \ , \ {u_1}_\perp = -{u_2}_\perp =-u\sin\theta \ .
\label{projection}
\end{equation}

The impulses relates the kinematic variables before and after the collision:
\begin{equation}
\begin{split}
{u_1'}_{\parallel} = {u_1}_{\parallel} + J_{\parallel}/M \ , \ {u_2'}_{\parallel} = {u_2}_{\parallel} - J_{\parallel}/M \ , &
\\
{u_1'}_{\perp} = {u_1}_{\perp} - J_{\perp}/M \ , \ {u_2'}_{\perp} = {u_2}_{\perp} + J_{\perp}/M \ , &
\\
\omega_1' = \omega_1 - J_{\parallel} R/I \ , \ \omega_2' = \omega_2 - J_{\parallel} R/I \ . &
\end{split}
\label{kinematic}
\end{equation}
As a sanity check, by direct substitution of Eq. \eqref{kinematic} one can show that the total angular momentum is conserved $L_{before} = L_{after}$ for all $\left( J_{\parallel}, J_{\perp} \right)$:
\begin{equation}
\begin{split}
L_{before} = I \omega_1 + I \omega_2 + M \left(\frac{b}2\right) u_1 - M \left(\frac{b}2\right) u_2 = I \omega_1 + I \omega_2 + M R {u_1}_{\parallel} - M R {u_2}_{\parallel} \ , &
\\
L_{after}  = I \omega_1' + I \omega_2' + M R {u_1'}_{\parallel} - M R {u_2'}_{\parallel} \ . &
\end{split}
\end{equation}
Since the spinners are gears with teeth, the no-sliding condition at their contact points in the $\hat{\parallel}$-direction is enforced right after the collision:
\begin{equation}
{u_1'}_{\parallel} - R\omega_1' = {u_2'}_{\parallel} + R\omega_2' \ . \label{no_sliding}
\end{equation}
Plug Eq. \eqref{projection}, Eq. \eqref{kinematic} into Eq. \eqref{no_sliding}, we can solve for $J_\parallel$:
\begin{equation}
\begin{split}
\left(u\cos\theta + J_{\parallel}/M \right) - R\left( \omega_1 -J_{\parallel}R/I \right) = \left( -u\cos\theta - J_{\parallel}/M \right)+R\left( \omega_2 - J_{\parallel}R/I\right) & 
\\
\Rightarrow \ J_{\parallel} = \frac{-2Mu\cos\theta + MR(\omega_1 + \omega_2)}{2\left( 1 + MR^2/I\right)} \ , &
\end{split}
\label{J_parallel_solve}
\end{equation}
and thus obtain the relation between angular velocities before and after the collision:
\begin{equation}
\begin{split}
\omega_1' = \omega_1 - \frac{J_{\parallel}R}I = \omega_1 - \frac{MR^2}{2\left( I + MR^2 \right)} \left( \omega_1 + \omega_2 \right) +\frac{2MRu\cos\theta }{2\left( I + MR^2 \right)} \ \xrightarrow[-2R < b < +2R]{\text{averaging over } b} \ \omega_1' - \omega_1 = - \frac{\omega_1 + \omega_2}{2\left( 1 + I/MR^2 \right)} \ , &
\\
\omega_2' = \omega_2 - \frac{J_{\parallel}R}I = \omega_2 - \frac{MR^2}{2\left( I + MR^2 \right)} \left( \omega_2 + \omega_1 \right) + \frac{2MRu\cos\theta }{2\left( I + MR^2 \right)} \ \xrightarrow[-2R < b < +2R]{\text{averaging over } b} \ \omega_2' - \omega_2 = - \frac{\omega_1 + \omega_2}{2\left( 1 + I/MR^2 \right)} \ ,
\end{split}
\end{equation}
Define the inertial parameter $\beta=1/2(1+I/MR^2)<1/2$, then after averaging over the impact parameter $b$ we get the simplification:
\begin{equation}
\omega_1' - \omega_1 = \omega_2' - \omega_2 = -\beta \left( \omega_1 + \omega_2 \right) \ .
\label{alphabet_of_collision}
\end{equation}
We will use this kinematic relationship to study the behavior of a many-spinner ensemble with two opposite-chirality ($+$ for counter-clockwise spinners and $-$ for clockwise spinners).

\subsubsection{B. The Emergence of Collective Spin-Pumping}

\begin{figure}[!bth]
\centering
\includegraphics[width=0.7\textwidth]{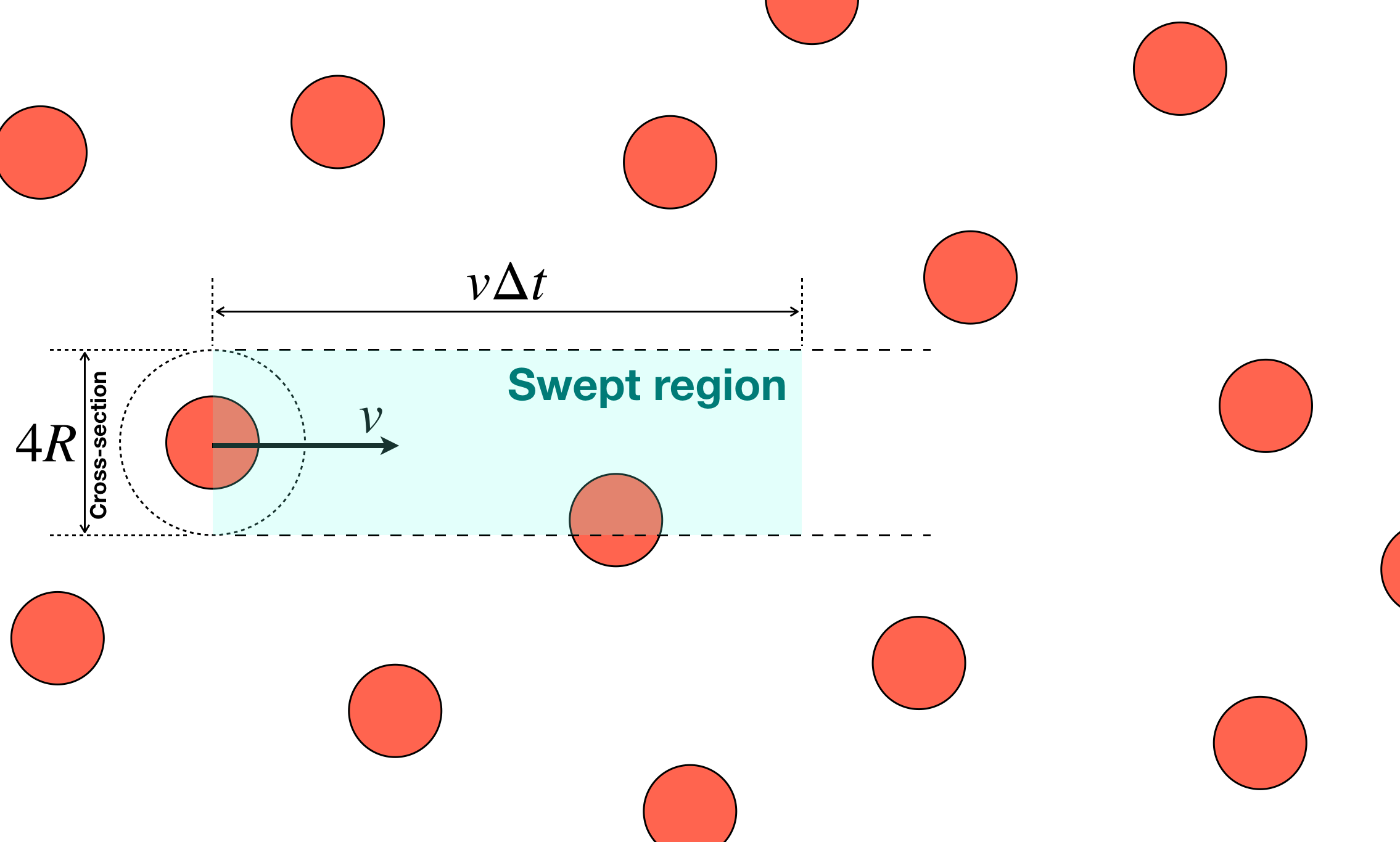}
\caption{During the time $\Delta t$, on average a spinner sweeps a region of area $4R\langle v \rangle \Delta t$.}
\label{figS04}
\end{figure}

Consider $N=N_++N_-$ total number of spinners with $N_+$ counter-clockwise spinners and $N_-$ clockwise spinners on an arena of area size $A$ and perimeter length $P$, for each spinner the available area of the other $N-1$ spinners is about $(A-PR)- (2R)^2$ since their centers cannot get closer than $2R$. If the average velocity is $\langle v \rangle$ then during the time $\Delta t$ a spinner can collide with any spinner inside a swept region of area $4R \langle v \rangle \Delta t$ (see Fig. \ref{figS04}). 

The characteristic collision time $\tau$ can be estimated by associating the time scale $\Delta t = \tau$ for the expectation of encountering $1$ other spinner in the swept region:
\begin{equation}
4R\langle v \rangle \Delta t \ \times \ \frac{N-1}{(A-PR) - \pi (2R)^2}  \Bigg|_{\Delta t = \tau}\sim 1 \ \ \Rightarrow \ \ \tau = \mathcal{O}(1) \ \times \ \left(\frac{A-PR-4\pi R^2}{4R \langle v \rangle} \right) \frac1{N-1} \ .
\end{equation}
Due to relative translational motion between spinners, the coefficient $\mathcal{O}(1)$ is roughly $\sim 1/\sqrt{2}$ \cite{kardar2007statistical}. We can lump the pre-factor together into $\epsilon \sim (A-PR-4\pi R^2)/4R\langle v \rangle$, so that $\tau = \epsilon/(N-1)$. Note that, in general, for $\tau$ as a function of density, even for the simplest cases of a two-dimensional hard-disk gas, the dependency is complicated. We refer to the following paper \cite{taloni2015collisional} for a more thorough and numerical calculation.

\begin{figure}[!htb]
\centering
\includegraphics[width=0.8\textwidth]{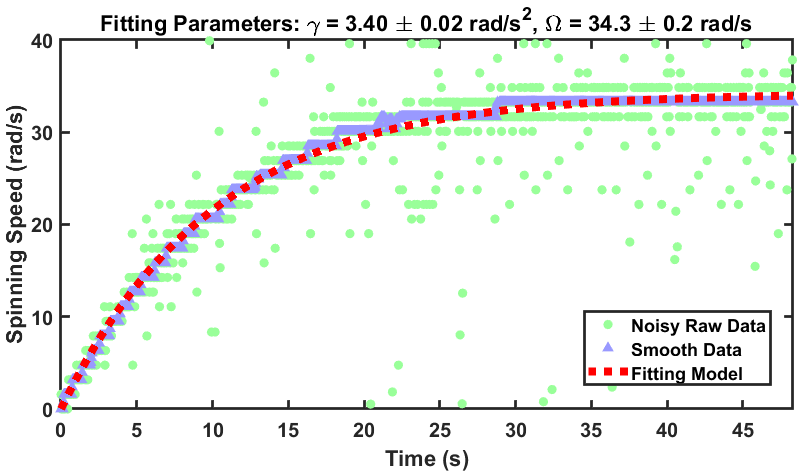}
\caption{We investigate how a single isolated spinner floating on the air-bed accelerate its spinning speed. Here we fit the model described by Eq. \eqref{Gamma_omega} with the data (the raw data was obtained at framerate 30 fps, the smooth data comes from moving median inside a 2-seconds window). The best-fit captures the observation nicely.}
\label{figS02}
\end{figure}

Now let us make a mathematical estimation for the average value of the spinning velocity $\langle \omega_\pm \rangle$:
\begin{equation}
 \frac{d}{dt} \langle \omega_\pm \rangle \approx
 \frac{N_\pm-1}{N-1} \Big[-\frac{\beta \left( \langle \omega_\pm \rangle + \langle \omega_\pm \rangle \right)}{\tau}  \Big] + \frac{N_\mp}{N-1} \Big[ - \frac{\beta \left( \langle \omega_\pm \rangle + \langle \omega_\mp \rangle \right)}{\tau}  \Big] \pm \frac{\int \Gamma_{\omega_\pm(t)} dt}{\tau} \Bigg|_{\langle \omega_{\pm} \rangle}\ ,
 \label{change_of_angular_vec}
\end{equation}
where on the right side the first term represents the change due to a collision between spinners with the same chirality, the second term represents the change due to a collision between spinners with the opposite chirality, as followed from Eq. \eqref{alphabet_of_collision}. The third term represents the average change of angular velocity between consecutive collisions. The angular acceleration $\pm\Gamma_{\omega_\pm}$ is a function of $\omega_\pm$, in general can be can be approximated by a Taylor's expansion as followed \cite{taylor1717methodus,thomas1968calculus}:
\begin{equation}
\pm \Gamma_{\omega_\pm} = \Gamma_\pm^{(0)} + \Gamma_\pm^{(1)} \omega_\pm + \mathcal{O}(\omega_\pm^2) \ .
\label{Gamma_omega_Taylor}
\end{equation}
Physically, we can assume that the air-blowers generate a constant torque contribution and there is a drag effect which is proportional to the spinning speed. Match with the description in Eq. \eqref{Gamma_omega_Taylor}, $\Gamma_\pm^{(0)}=\pm \gamma$ is set by that torque and $\Gamma_\pm^{(0)}=-\eta_{rot}$ is set by that drag. Define $\Omega = \gamma/\eta_{rot}$, we get:
\begin{equation}
\pm \Gamma_{\omega_\pm} = \pm \gamma - \eta_{rot} \omega_\pm = \pm \gamma \left(1 \mp \frac{\omega_\pm}{ \Omega} \right) \ ,
\label{Gamma_omega}
\end{equation}
in which $\Omega$ is the maximum possible angular-velocity of a single isolated spinners plays a similar role to that of the carrying capacity in growth dynamics \cite{phan2021doesn}. This model turns out to be in great agreement with how a single isolated aerial spinner accelerates its rotational motion (see Fig. \ref{figS02}).

This linearity in $\Gamma_\omega$ simplifies Eq. \eqref{change_of_angular_vec}:
\begin{equation}
\begin{split}
\frac{\int \Gamma_{\omega_\pm (t)} dt }{\tau} = \frac{\int \gamma \left( 1 \mp \omega_\pm (t)/\Omega \right) dt}{\tau} = \gamma \left[ 1 \mp \frac{\int \omega_\pm (t) dt/\tau }{\Omega} \right] = \gamma \left( 1 \mp \frac{\langle \omega_\pm \rangle }{\Omega} \right) = \Gamma_{\langle \omega_\pm \rangle} &
\\
\Rightarrow \ \frac{d}{dt} \langle \omega_\pm \rangle \approx
 \frac{N_\pm-1}{N-1} \Big[-\frac{\beta \left( \langle \omega_\pm \rangle + \langle \omega_\pm \rangle \right)}{\tau} \Big] + \frac{N_\mp}{N-1} \Big[ - \frac{\beta \left( \langle \omega_\pm \rangle + \langle \omega_\mp \rangle \right)}{\tau} \Big] \pm \Gamma_{\langle \omega_{\pm}\rangle} \ . &
 \label{change_of_angular_vec_simp}
 \end{split}
\end{equation}
At the steady state, $d\langle \omega_\pm \rangle / dt = 0$, therefore we can solve Eq. \eqref{change_of_angular_vec_simp} which is now just an algebraic equation to get:
\begin{equation}
\langle \omega_\pm \rangle = \pm \Omega \ \times \  \frac{\alpha \left( \alpha + 3N - 2 - 2N_\pm \right)}{(\alpha + 2N-2)(\alpha + N -2)}\ ,
\label{spin_pump_eq}
\end{equation}
where $\alpha = \gamma \epsilon/\beta\Omega$ depends on the spinner translational locomotion and inertial properties, and the air-blowing strength. Use $\gamma \sim 3$rad/s$^2$ and $\Omega \sim 30$rad/s as we find from Fig. \ref{figS02}, $\beta \sim 3/8$ (from the spinner inertial measured in Section I) and $\epsilon \sim 25$s (we get this from the average translational velocity $\langle v \rangle \sim 7$cm/s, the spinner radius $R=3.5$cm, the arena has area size $A\sim55$cm$\times 75$cm and perimeter length $P \sim 2(55+75)=260$cm, and $\mathcal{O}(1)\sim 1/\sqrt{2}$), we can make an estimation that $\alpha \sim 7$.

\begin{figure}[!htb]
\centering
\includegraphics[width=\textwidth]{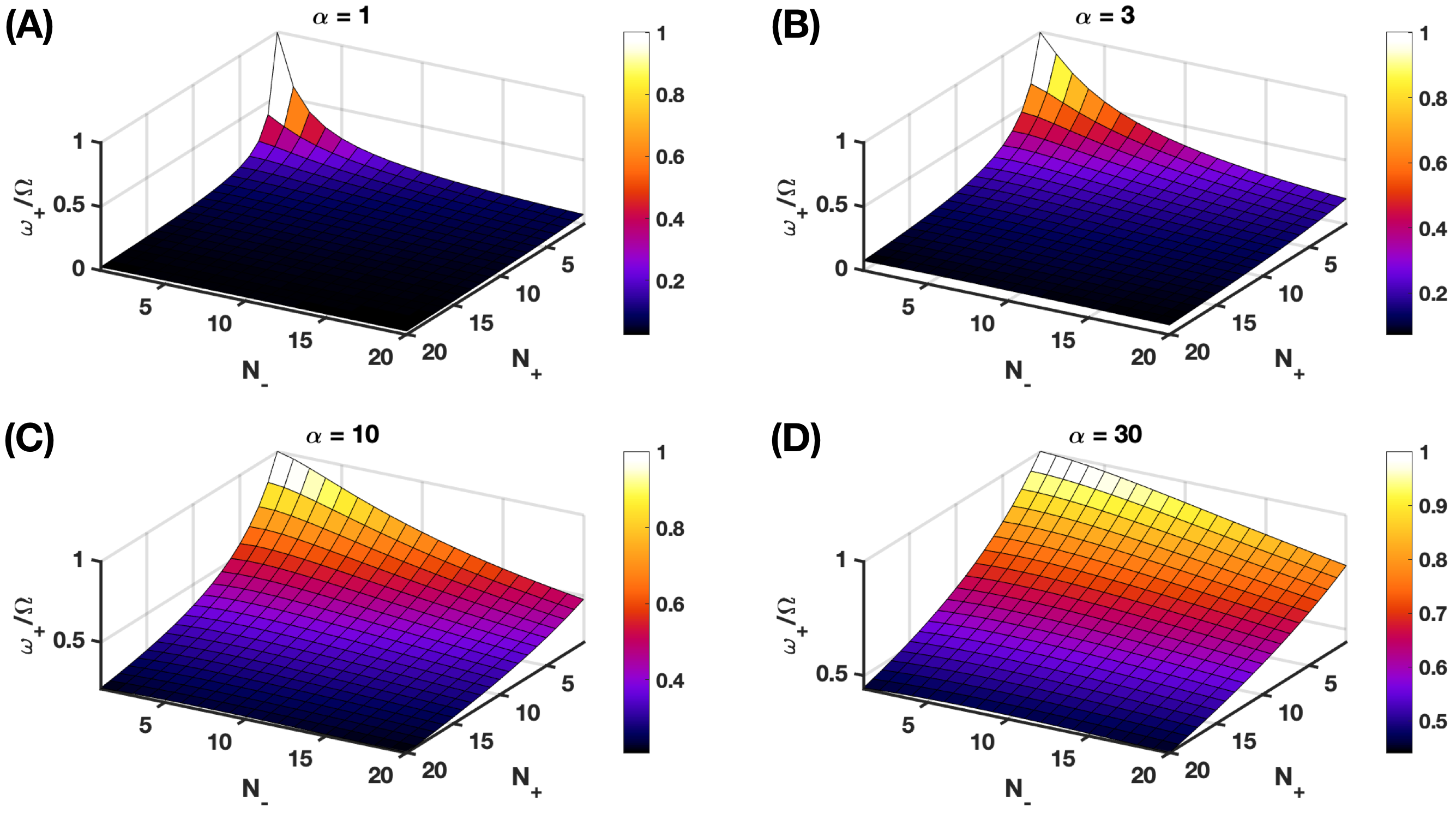}
\caption{The average spinning velocity $\langle \omega_+ \rangle$ in the unit of $\Omega$ as a function of $N_+$ and $N_-$ using Eq. \eqref{spin_pump_eq}, for different values of the parameter $\alpha$: (A) $\alpha=1$. (B) $\alpha=3$. (C) $\alpha=10$. (D) $\alpha=30$.}
\label{figS03}
\end{figure}

Note that Eq. \eqref{spin_pump_eq} applicable only with $N>1$ and for $\langle \omega_\pm \rangle $ when $N_\pm > 0$. When $N_+=1$ and $N_-=1$ we have $\langle \omega_\pm \rangle = \pm \Omega$, and for a fixed value of $N$ then $\langle \omega_\pm \rangle $ monotonically decreases as $N_\pm$ increases: the minority will always spin faster than the majority! This finding is presented in Fig. \ref{figS03}. When there is only one spinner we have $|\langle \omega \rangle| = \Omega$. For a pure population of, without loss of generality, $N_+=N>1$ counter-clockwise spinners, then the average angular velocity at the steady state is:
\begin{equation}
\langle \omega_+ \rangle = + \Omega \ \times \ \frac{\alpha}{\alpha+2N-2} \ .
\label{single_species}
\end{equation}
This means the more spinners the slower they can spin on average.

In the limit $N \rightarrow \infty$, define $n_+=N_+/N$ to be a fraction of counter-clockwise spinners, Eq. \eqref{spin_pump_eq} simplifies:
\begin{equation}
\langle \omega_\pm \rangle = \pm \Omega \ \times \ \frac{\alpha \left(3 - 2n_+\right)}{2N} \ .
\label{average_rot_velocity_simp}
\end{equation}

\subsection{IV. Average Rotational and Translational Energy of Spinners} 

\subsubsection{A. Rotational Energy}

\begin{figure}[ht!] 
\centering
\includegraphics[width=0.8\textwidth]{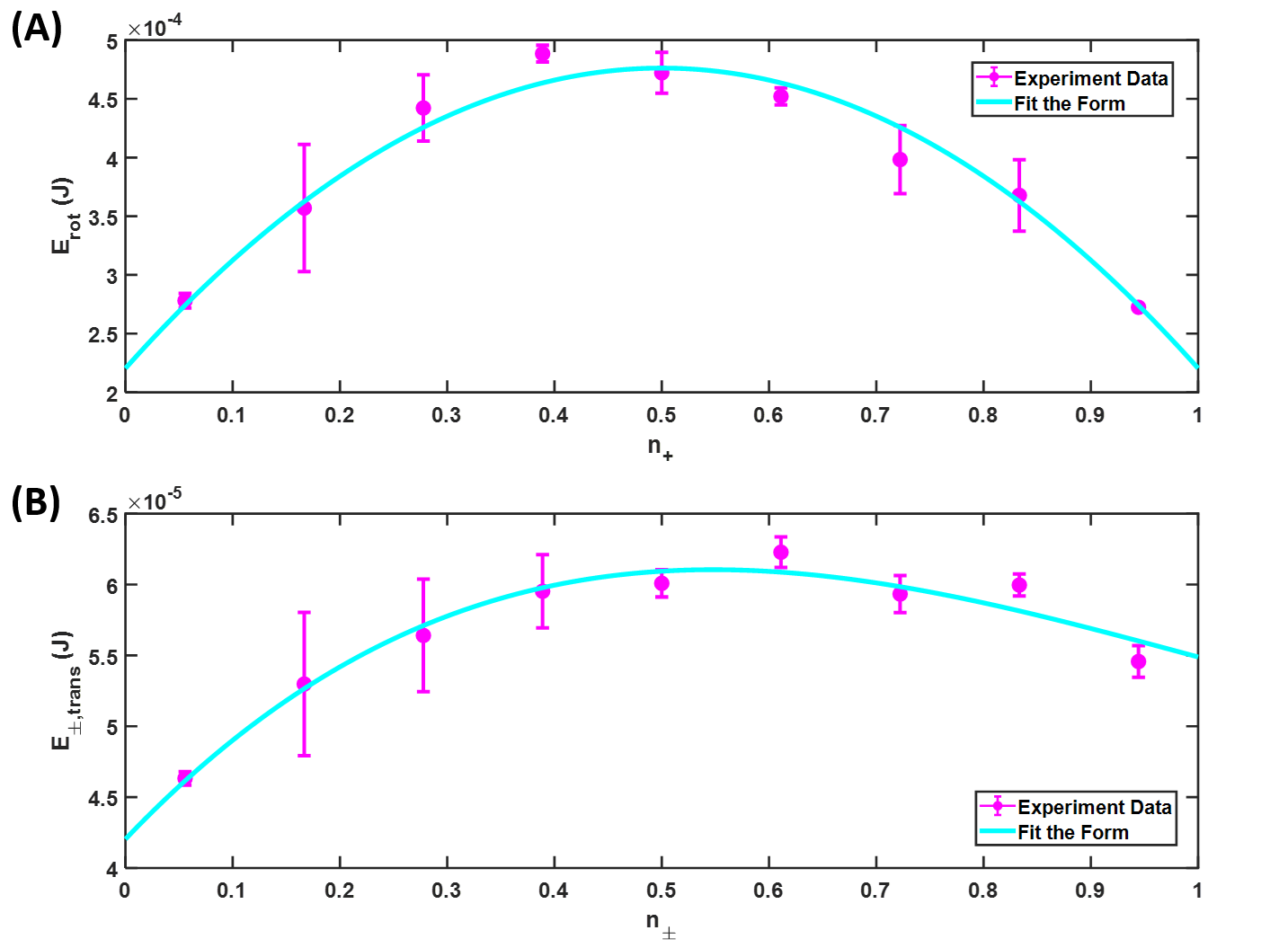}
\caption{(A) Rotational energy $E_{rot}$ as a function of the fraction $n_+$ of the spinner species, fitting with the form Eq. \eqref{tot_rotational_energy} up to a positive-constant. (B) Translational energy $E_{\pm,trans}$ as a function of the fraction $n_\pm$ of the spinner species, fitting with the form Eq. \eqref{average_trans_energy} up to a positive-constant.}
\label{figS06}
\end{figure}

We can also make another crude estimation, for the average rotational energy $E_{rot}$ of all spinners, using Eq. \eqref{spin_pump_eq} and take the average rotational energy of each species ($\pm$) of spinners to be $E_{\pm,rot}=I\langle \omega_\pm \rangle^2/2$:
\begin{equation}
\begin{split}
E_{rot} = \frac{N_+ E_{+,rot} + N_- E_{-,rot}}{N} = \frac12 I\Omega^2 \ \times \ \frac{\alpha^2 \left[ N_+ \left( \alpha + 3N - 2 - 2N_+\right)^2 + N_- \left( \alpha + 3N - 2 - 2N_-\right)^2 \right]}{N(\alpha + 2N - 2)^2 (\alpha + N - 2)^2}
\\
= \frac12 I\Omega^2 \ \times \ \frac{\alpha^2 \left[ N(\alpha + N - 2)^2 + 4N(2\alpha + 3N - 4) N_+ - 4 (2\alpha + 3N - 4) N_+^2 \right]}{N(\alpha + 2N - 2)^2 (\alpha + N - 2)^2}
\\
= \frac12 I\Omega^2 \ \times \ \frac{\alpha^2 \left[ N(\alpha + N - 2)^2 + (2\alpha + 3N - 4) N^2 - 4 (2\alpha + 3N - 4) (N/2 - N_+)^2 \right]}{N(\alpha + 2N - 2)^2 (\alpha + N - 2)^2} \ .
\end{split}
\label{tot_rotational_energy}
\end{equation}
Here we expand $E_{rot}$ as a sum series of $N_+$ (without loss of generality) and group then in a way that symmetry around $N/2$ (due to the interchangeable $N_+ \leftrightarrow N_-$). As a function of $N_+$ for a fixed value of $N$, we find that $E_{rot}$ is an inverted parabola centered at $N_+ = N/2$. This is consistent with what we have found in Fig. 2B of the main manuscript. In the limit $N \rightarrow \infty$, Eq. \eqref{tot_rotational_energy} simplifies:
\begin{equation}
E_{rot} = \frac12 I\Omega^2 \ \times \ \frac{\alpha^2\left[1 - 3\left(1/2-n_+ \right)^2\right]}{N^2} \ .
\label{average_rot_all}
\end{equation}
Up to a positive-constant energy shift, the form of Eq. \eqref{tot_rotational_energy} captures our observation quite nicely (see Fig. \ref{figS06}A).

\subsection{B. Translational Energy}
We will use some qualitative argument to explain why maximum translational energy $E_{\pm,trans}$ of each species are somewhere between $n_\pm \in [1/2,1]$ as shown in Fig. 2B of the main manuscript. In this system, the incoming energy is mostly from the blowers that make the spinners rotate and the air flows that drive the spinners moving. Also collisions between spinners of the same species turn rotational energy into translational energy. Thus the higher the rotation energy $E_{\pm,rot}$ and the more collisions between spinners of the same species $\propto n_\pm$, the bigger the translational energy $E_{\pm,trans}$ can be. Utilize the simplification Eq. \eqref{average_rot_velocity_simp}, we obtain:
\begin{equation}
E_{\pm,trans} \propto \left( E_{rot} + E_{air} \right) n_\pm \propto \left[(3-2n_\pm)^2 + \mathcal{E}_{air} \right] n_\pm  \ ,
\label{average_trans_energy}
\end{equation}
where $\mathcal{E}_{air}>0$ are is the possible contributions from the air flows. The maximum of this function in the range of $n_\pm \in [0,1]$ should be inside $n_\pm \in [0,1]$ for $\mathcal{E}_{air}\leq 3$:
\begin{equation}
n_{\pm}\Big|_{\max(E_{\pm,trans})} = 1 - \frac12 \sqrt{1-\frac{\mathcal{E}_{air}}3} \ ,
\end{equation}
and right at the upper-limit value $n_{\pm}=1$ for $\mathcal{E}_{air}>3$. Up to a positive-constant energy shift, the form of Eq. \eqref{average_trans_energy} captures our observation quite nicely (see Fig. \ref{figS06}B).

\subsection{V. The Damped Oscillations of Mixing Entropy}

\begin{figure}[!htb]
\centering
\includegraphics[width=\textwidth]{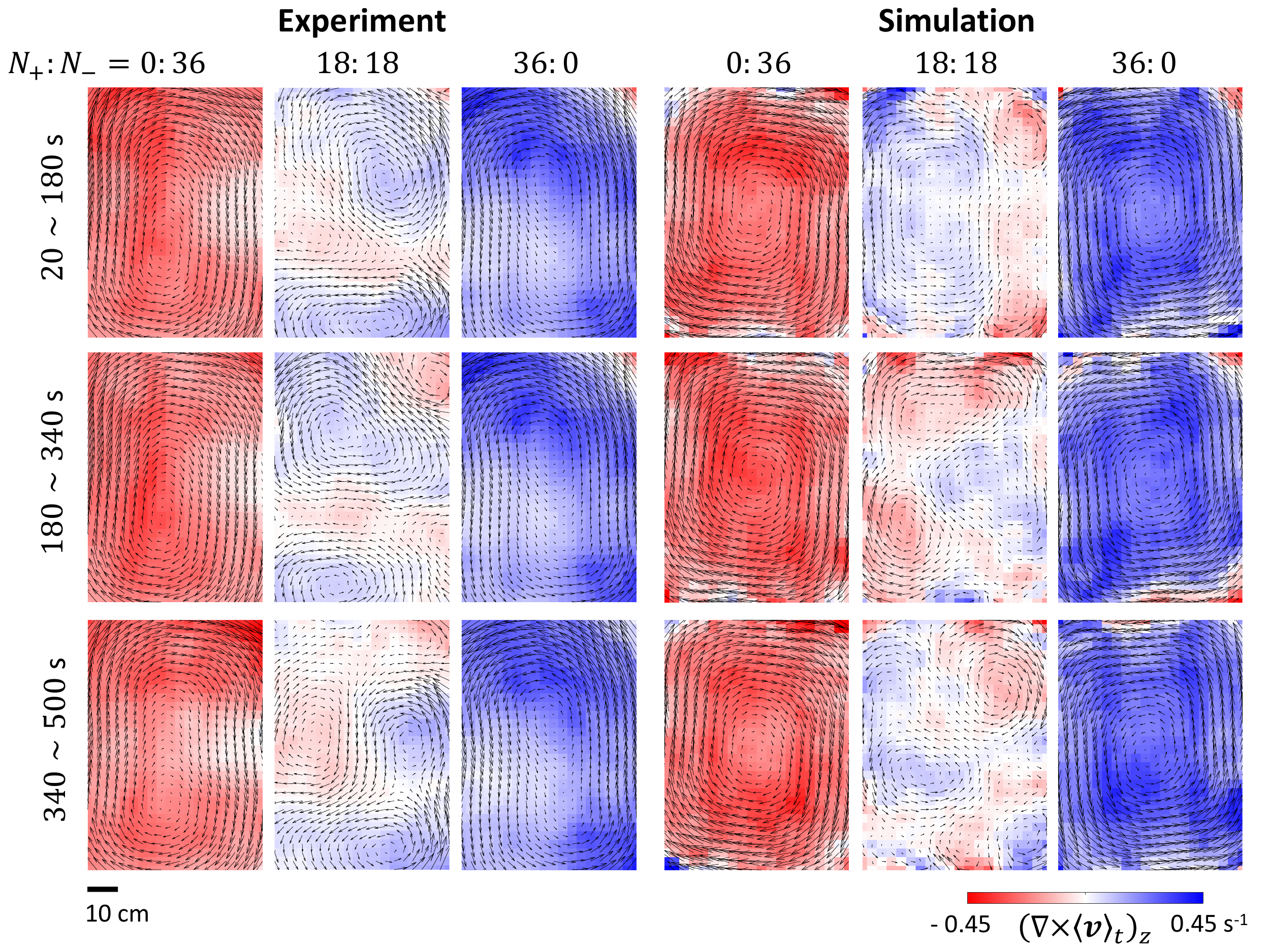} 
\caption{The evolution of the time-window-averaged velocity profile for 36-spinner collectives in which $N_+:N_-$ are $36:0$, $18:18$ and $0:36$. The color shows the vorticity ($\zeta = (\nabla \times \langle \mathbf{v}\rangle)_z$). From the time evolution, we can see there are multiple vortices in the even-mixture case and the vortices are motile in both simulation and experiment. To count the number of vortices (Fig.3F in the main text), we use the bwconncomp function in MATLAB to find the number of connected components for $\zeta>0$ and $\zeta<0$. Vortices larger than $3\%$ of the arena sizes are considered.}
\label{figS09}
\end{figure}

From the experiments, we can see that at a high enough population of spinners, a topological current can emerge on the outer-most layer while a jammed inner core can be formed. Consider $N=36$, we observe that for $N_+:N_-=36:0$ the edge current circulates counter-clockwise, for $N_+:N_-=0:36$ the edge current circulates clockwise, and for $N_+:N_-=18:18$ there is no clear sign for an edge current (see Fig. \ref{figS09}). At lower-density, there is no edge current and the inner core is not jammed. While the following uses a graphic guidance where there is a clear distinction between a rotating outer layer and static inner core, in real situation where such distinction is more ambiguous, the inner core generalizes to an idle portion of spinner and the outer flow generalizes to the circulating spinners. We posit this simplified model still qualitatively captures the phenomenon.

\begin{figure}[!htb]
\centering
\includegraphics[width=0.8\textwidth]{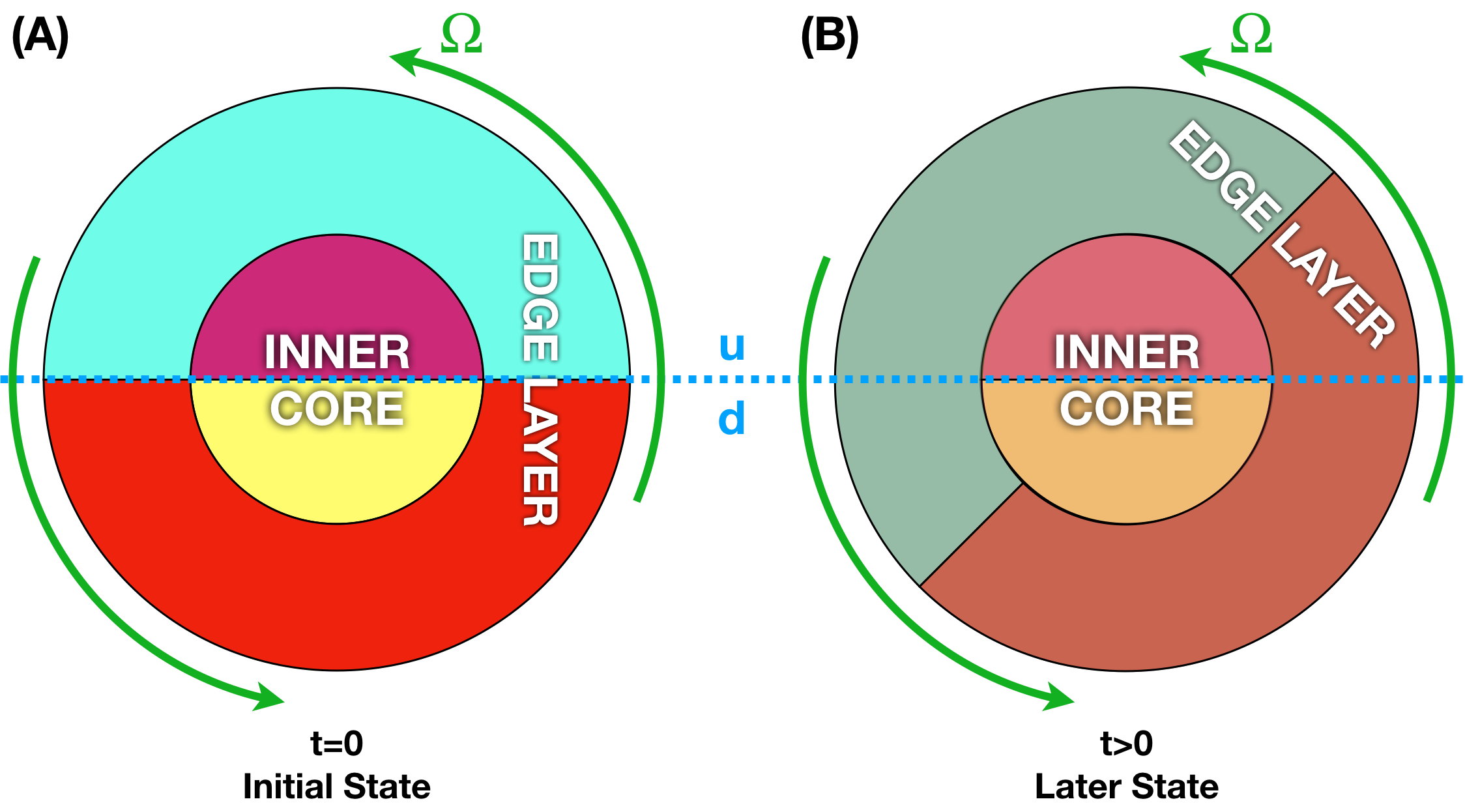}
\caption{The emergent of a topological edge current can explain what is going on with mixing entropy and how recurrences can happen.}
\label{figS05}
\end{figure}

Here we show how mixing entropy can oscillate due the emergent of a topological edge current at high-density of single species spiners. Say, in the beginning $t=0$ (see Fig. \ref{figS05}A), the upper-half and the lower-half of the arena has equal number of spinners $N/2$. There are $O^u_u(0)=O^d_d(0)=\nu N/2$ in the upper-half and lower-half of the edge layer (which corresponds to the fast circulating boundary flow) and $I^u_u(0)=I^d_d(0)=(1-\nu) N/2$ in the upper-half and lower-half of the inner core, where $\nu<1$ represents the fraction of the spinners on the edge layer. The low-density limit corresponds to $\nu \rightarrow 0$. 

First, let's consider the edge layer does not circulate around the inner core. To describe the mixing phenomenon, we will assume the following simple dependency:
\begin{equation}
\begin{split}
\tilde{O}^u_u(t) = O^u_u(0) \left(\frac{1+e^{-t/T_O}}2 \right) \ , \ \tilde{O}^d_d(t) = O^d_d(0) \left( \frac{1+e^{-t/T_O}}2 \right) \ ,
\\
\tilde{O}^d_u(t) = O^u_u(0) - \tilde{O}^u_u(t) \ , \ \tilde{O}^u_d(t) = O^d_d(0) - \tilde{O}^d_d(t)  \ ,
\end{split}
\end{equation}
and:
\begin{equation}
\begin{split}
\tilde{I}^u_u(t) = I^u_u(0) \left(\frac{1+e^{-t/T_I}}2 \right) \ , \ \tilde{I}^d_d(t) =  I^d_d(0) \left( \frac{1+e^{-t/T_I}}2 \right) \ ,
\\
\tilde{I}^d_u(t) = I^u_u(0) - \tilde{I}^u_u(t) \ , \ \tilde{I}^u_d(t) = I^d_d(0) - \tilde{I}^d_d(t)  \ .
\end{split}
\end{equation}
These equations only have mixing in the edge layer only and the inner core only but no exchange of spinners between these. Of course reality, there should also be mixing between the edge layer and the inner core too, but the results that we see with our simplification will not be different (and the advantage is that the math becomes much more tractable).

\begin{figure}[!htb]
\centering
\includegraphics[width=0.8\textwidth]{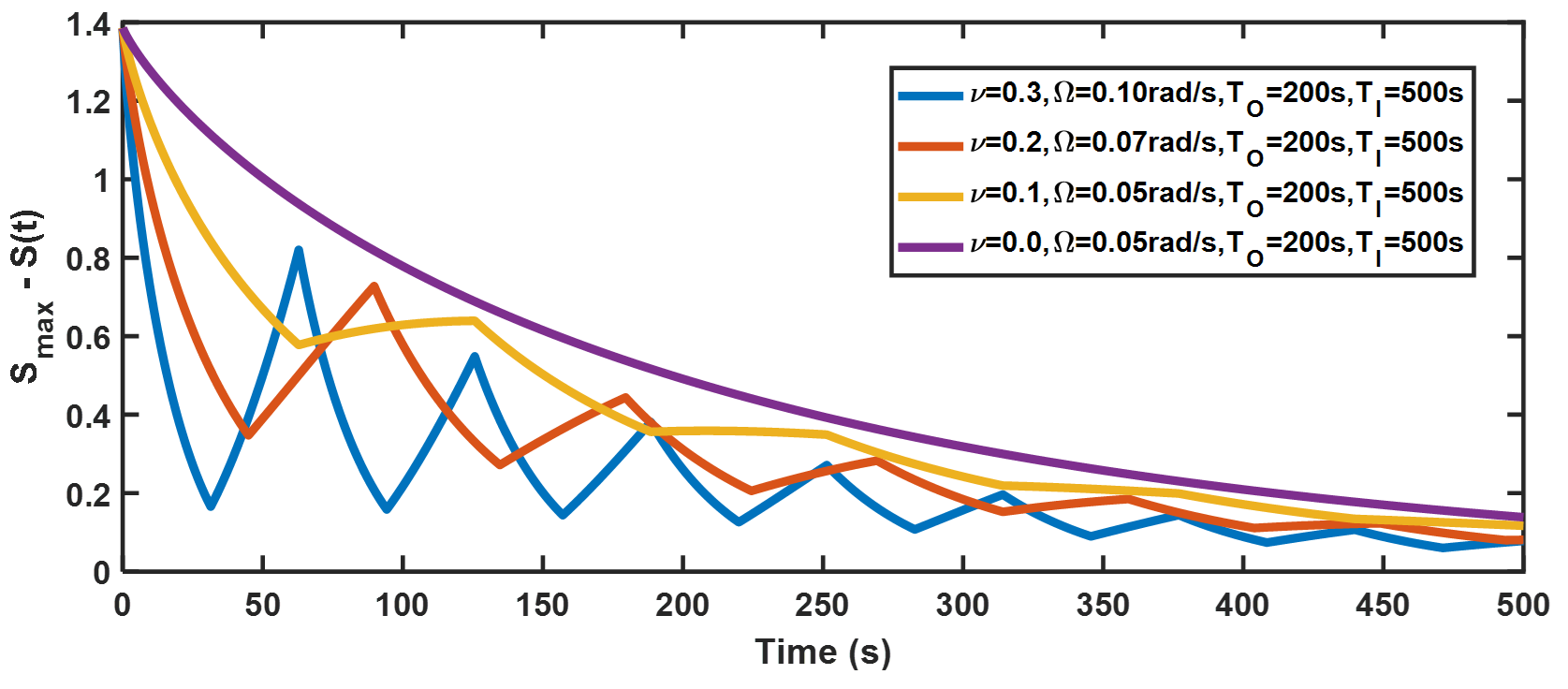}
\caption{The mixing entropy $S(t)$ as given in Eq. \eqref{Mixing_Entropy_Model} exhibits a damped oscillating behavior.}
\label{figS08}
\end{figure}

For the edge layer circulates around the inner core with angular velocity $\Omega$ (see Fig. \ref{figS05}B), we have:
\begin{equation}
\begin{split}
O^u_u(t) = \tilde{O}^u_u(t)\Delta \left(\frac{\pi - \Omega t}{2\pi} \right) + \tilde{O}^d_u(t)\Delta \left(\frac{\Omega t}{2\pi} \right)  \ ,
\\
O^d_d(t) = \tilde{O}^d_d(t)\Delta \left(\frac{\pi - \Omega t}{2\pi} \right)  + \tilde{O}^u_d(t)\Delta \left(\frac{\Omega t}{2\pi} \right)  \ ,
\end{split}
\end{equation}
and:
\begin{equation}
I^u_u(t) = \tilde{I}^u_u(t)  \ , \ I^d_d(t) = \tilde{I}^d_d(t)  \ ,
\end{equation}
in which we use the triangle-wave function:
\begin{equation}
\Delta(\xi) = 2 \left|\xi - \left\lfloor \xi + \frac12 \right\rfloor \right| .
\end{equation}
The joint probabilities are given by:
\begin{equation}
p^u_u = \frac{O^u_u(t) + I^u_u(t)}{O^u_u(0) + I^u_u(0)} \ , \ p^d_d = \frac{O^d_d(t) + I^d_d(t)}{O^d_d(0) + I^d_d(0)} \ ,
\end{equation}
and the mixing entropy can be calculated as:
\begin{equation}
\begin{split}
S(t) &= - p^u_u \ln p^u_u -p^d_u \ln p^d_u - p^d_d \ln p^d_d - p^u_d \ln p^u_d
\\
&= - p^u_u \ln p^u_u -\left(1 - p^u_u \right) \ln \left(1 - p^u_u \right) - p^d_d \ln p^d_d - \left(1 - p^d_d \right) \ln \left(1 - p^d_d \right) \ .
\end{split}
\label{Mixing_Entropy_Model}
\end{equation}
We can plot $S(t)$ depends on the parameters $(\Omega,T_O,T_I)$ as shown in Fig. \ref{figS08}. While the function is not as smooth as seen in our experiments, this model still qualitatively captures the damping oscillation nonetheless. Moreover, if we change $\Delta(\cdot)$ to $\sin{(\cdot)}$, the evolution will become smoother and closer to the data in experiments due to the smear-out of the rotating spinner strip.

The recurrence time $T_{recc}$ is equal to half of the circulation time $T_{circ}=2\pi/\Omega$. From the experiments we see $T_{circ}\sim60$s, therefore we can estimate $T_{recc}\sim 30$s. This finding is indeed in agreement with the mixing entropy oscillation!

\newpage

\subsection{VI. Gears made of Gears}

\begin{figure}[!htb]
\centering
\includegraphics[width=\textwidth]{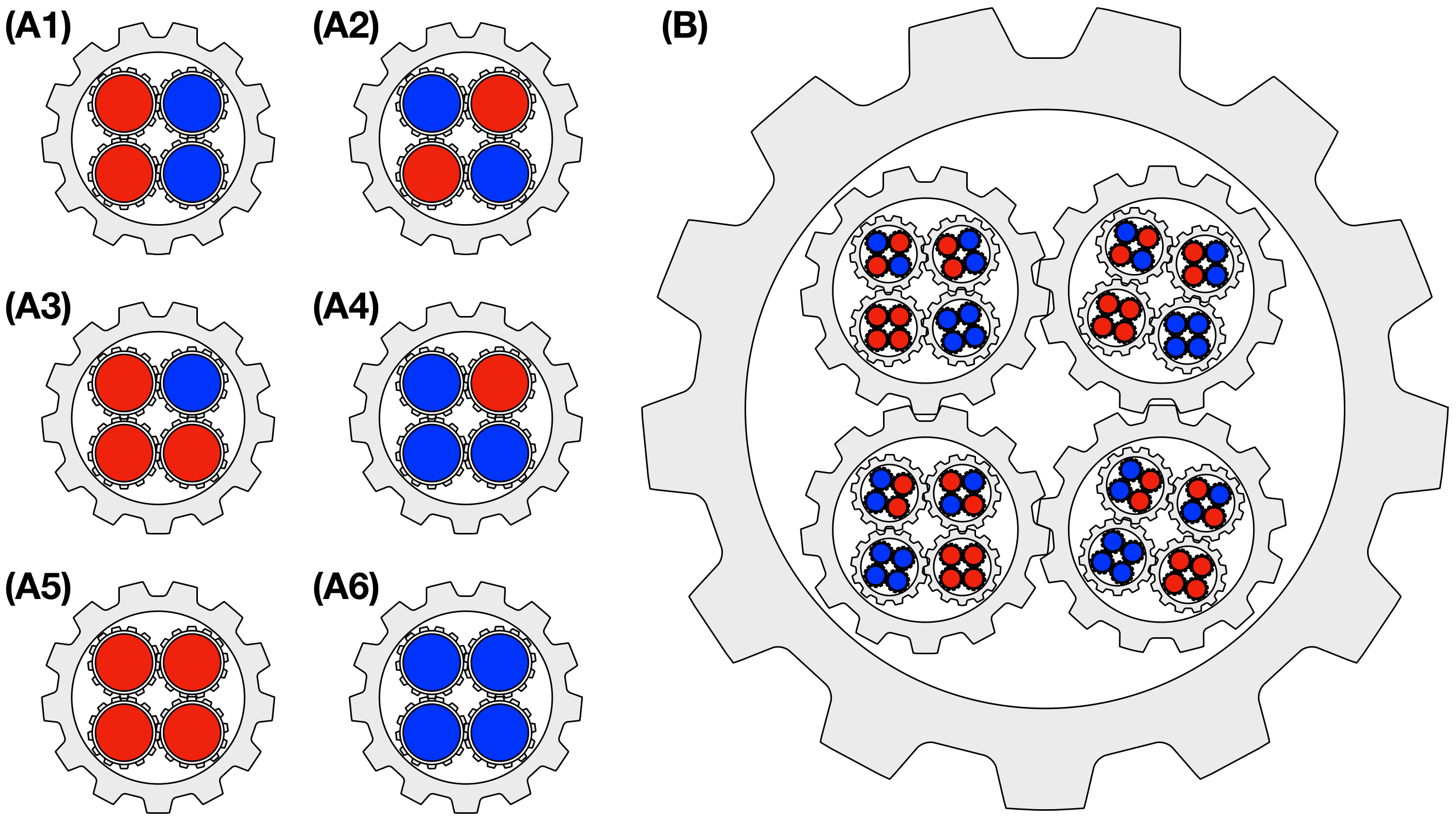}
\caption{(A1-6) Some examples of gears made of fundamental gears, which are spinners. Here we list all possible topological arrangement with four fundamental gears, and (A1-2) are the designs we investigate in the main manuscript. (B) A scale-free fractal gear, which is three-level above the fundamental gears, larger in size but same designing and operation principles.}
\label{figS10}
\end{figure}

Gears are toothed, mechanical transmission elements used to transfer motion and power between components of machines. They are ubiquitous, can be found within a large range of length-scales. Since the designing and operation principles of gears are independence of size, it is very convenient to use them for creating scale-free fractal structures in which low-level small gears drive higher-level big gears. We can use our spinners as the fundamental gears, put some of them inside a ring with teeth to get a structure which can also works like a gear (see Fig. \ref{figS10}A). Repeating this we can build Matryoshka superstructures of scale-free fractal gears (see Fig. \ref{figS10}B).

For 4 spinners geometrically confined by an outer gear which is a toothed ring as described in the main manuscripts, there are 6 possible arrangements (see Fig. \ref{figS10}A1-6) with spaces in between gears. It should be noted that, for simple design, the ring only has outer-teeth but no inner-teeth, which means the inside gears powered the outside gears via frictional couplings.

\begin{figure}[!htb]
\centering
\includegraphics[width=\textwidth]{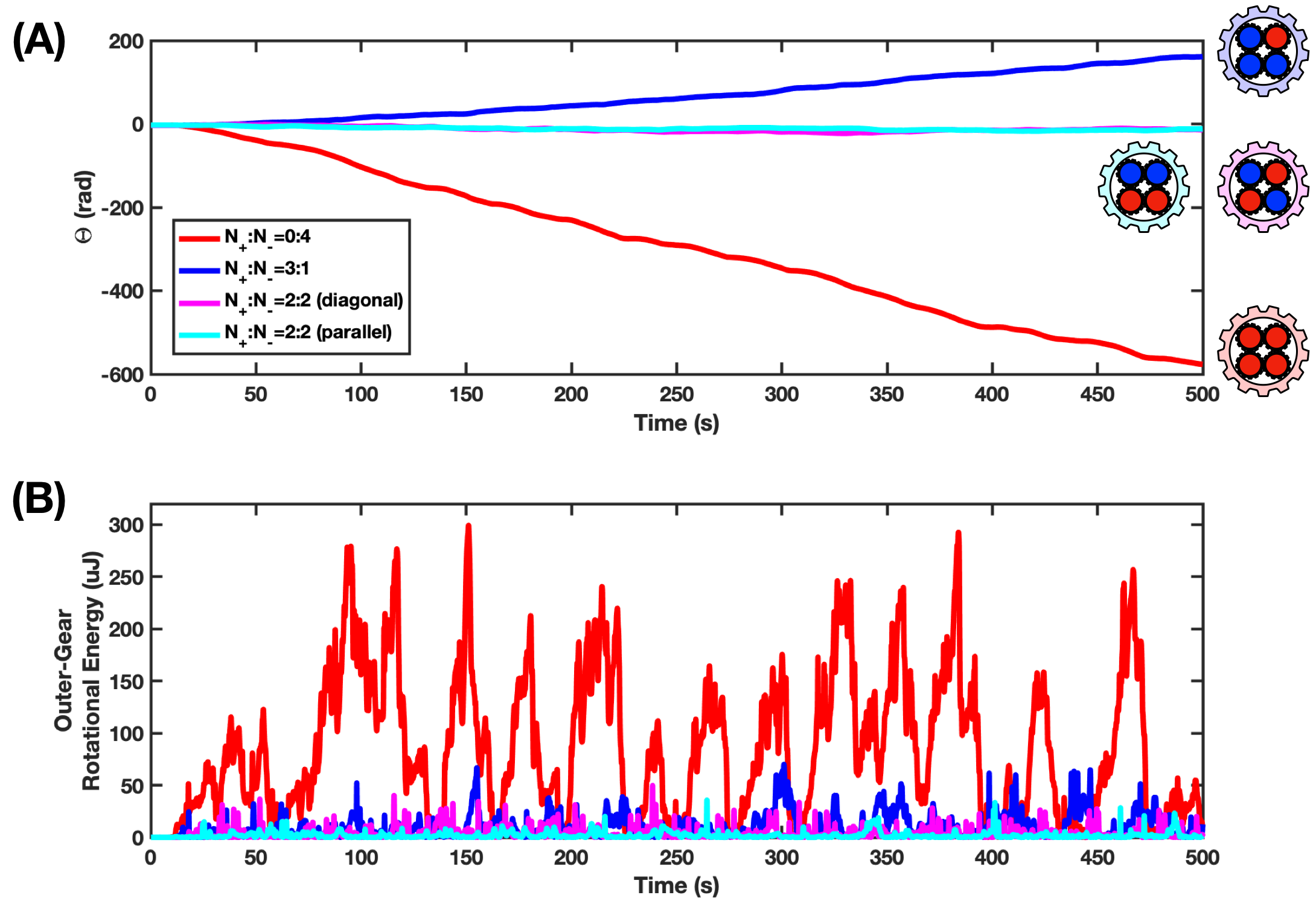}
\caption{(A) The rotation of the outer-gear, angle $\Theta$ is positive if it is turned counter-clockwise. (B) The rotational energy of the outer-gear, which exhibits an intermittency behavior for all arrangements of 4 spinners inside.}
\label{figS11}
\end{figure}

There are many possible configurations for the positions of spinners and for the locking-interaction between them. This degeneracy gives rise to spinning frustration -- there exists no steady state of spinning as these gears made of gears keeps jumping between different configurations which can be seen from the intermittency of their rotational motion. We call them the frustrated spinning states. We show that each frustrated spinning state corresponds to a rotational mode, with distinct average angular velocity $\langle \dot{\Theta} \rangle$ and rotational energy as shown in Fig. \ref{figS11}. The more homogeneous (single-species) the faster they can spin, and for equal number of spin-up and spin-down they pretty much cannot spin (due to symmetry cancelling out torque contributions). 

The special cases $N_+:N_-=2:2$, which is considered in the main manuscripts, has two possible topological arrangements (parallel or diagonal same-species). Even though they both does not spin very much on average, the average and intermittency of rotational energies are not the same.

\clearpage
\subsection{VII. Simulation} 

\subsubsection{A. Model}
\begin{figure}[!htb]
\centering
\includegraphics[width=0.8\textwidth]{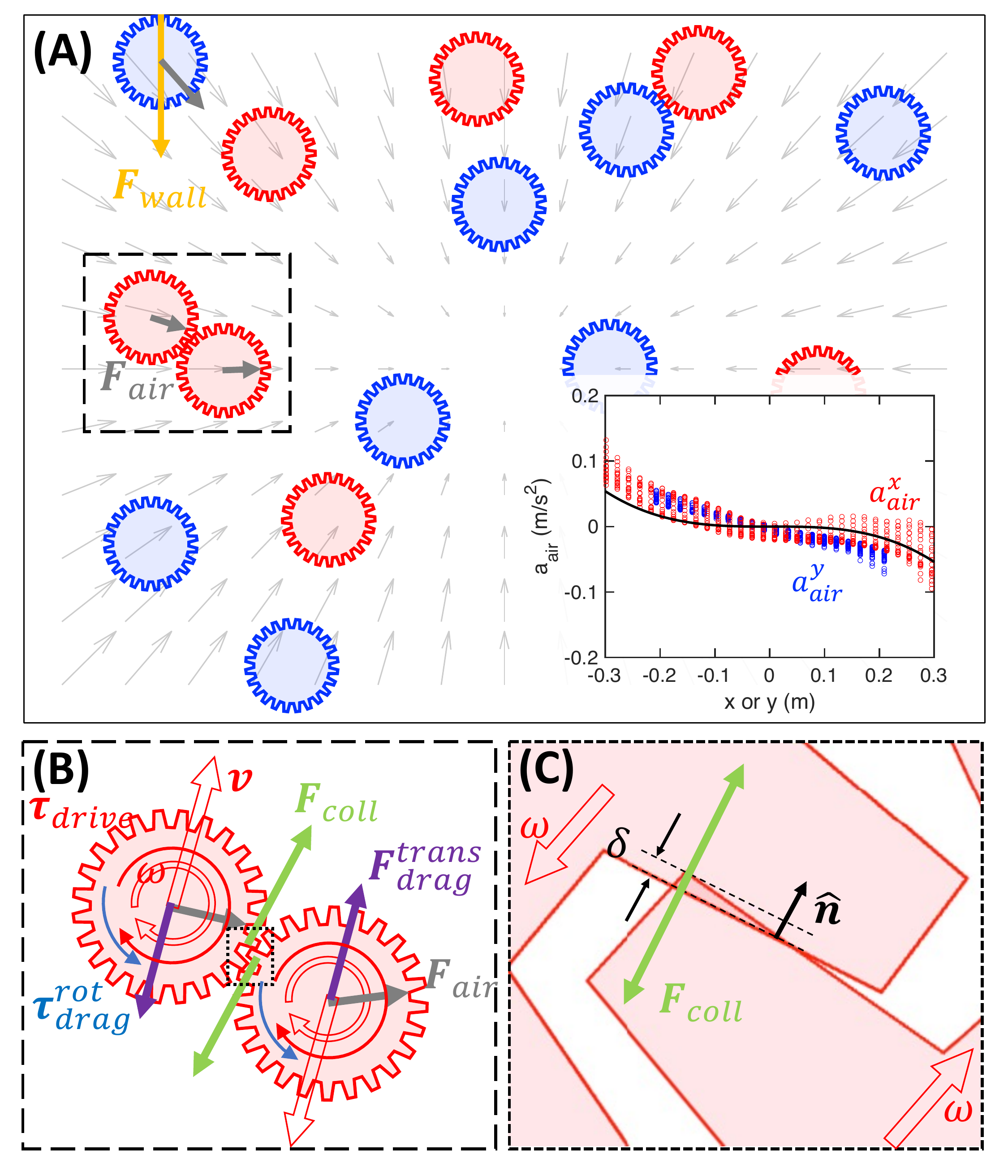}
\caption{\textbf{Simulation setup.} (A) Each spinner is subjected to the air current force $\mathbf{F}_{air}$ and the normal force from the boundary $\mathbf{F}_{wall}$. Inset: the components of $\mathbf{a}_{air}=\mathbf{F}_{air}/m$. The black cubic fit shows the air current force used in simulation. (B) Besides the collision force $\mathbf{F}_{coll}$ from another spinner, each spinner is also subjected to a translational drag force $\mathbf{F}_{drag}^{trans}$, which is antiparallel to the velocity $\mathbf{v}$. In the rotational direction, each spinner is subjected to a driving torque $\tau_{drive}$ from the blowers on the spinner and a rotational drag torque $\tau_{drag}^{rot}$. (C) In simulation, the collision force uses a spring-dash model where the collision force is composed of a restoring force proportional to the virtual overlap $\delta$ and a dissipation dash force proportional to both $\delta$ and the approaching velocity's projection in the normal direction $\hat{\mathbf{n}}$.}
\label{fig:simSetup}
\end{figure}

In the simulation, each spinner is subjected to the collision force from another spinner $\mathbf{F}_{coll}$, translational drag force $\mathbf{F}_{drag}^{trans}$, rotational driving torque $\tau_{drive}$ from the blowers on spinner, rotational drag torque $\tau_{drag}^{rot}$, air current flow $\mathbf{F}_{air}$, and collision force from the boundary $\mathbf{F}_{wall}$ (Fig.\ref{fig:simSetup}A,B). The parameters are determined from direct and indirect physical measurement as listed in the following table.

\begin{table}[!htb] 
    \centering
    \begin{tabular}{|clll|}
        \hline
        & \textbf{Description} & \textbf{Value} & \textbf{Reference} \\
        \hline\hline
        $m$ & Spinner mass & 0.025 kg & Direct measurement from a scale \\ 
        \hline
        $R_0$ & Spinner outer radius & 0.035 m & Direct measurement from a caliper \\
        \hline
        $R_i$ & Spinner inner radius & 0.030 m & Same as above\\
        \hline
        $I$ & Spinner moment of inertia & $1.01\times 10^{-5}$ kg m$^2$ & Sec. II A of this document\\
        \hline
        $\Omega$ & \makecell[l]{Saturated angular velocity\\ of the orbit} & 32 rad s$^{-1}$ & Fig.\ref{figS02}\\
        \hline
        $\gamma$ & Rotational driving acceleration & 3.4 rad s$^{-2}$ & Same as above \\
        \hline
        $\tau_{drive}$ & Rotational driving torque &  $3.4 \times 10^{-5}$ N m & $=I\gamma$ \\
        \hline
        $\eta$ & Translational drag coefficient & $1.6\times 10^{-3}$ kg s$^{-1}$ & $=m\cdot 6.5\times 10^{-2}$ s$^{-1}$. See Fig.\ref{figS07}D.\\
        \hline
        $\eta_{\varphi}$ & \makecell[l]{Rotational drag coefficient} &  $1.06\times 10^{-6}$ N m s & $=\tau_{drive}/\Omega$\\
        \hline
        $F_{air}$ & Air current force & & See the cubic fit in Fig.\ref{fig:simSetup}A inset.\\
        \hline
        $A_1$ & Half arena length & $0.38$ m & Direct measurement from a meter stick\\
        \hline
        $A_2$ & Half arena width & $0.28$ m & Same as above\\
        \hline
    \end{tabular}
    \caption{\textbf{List of parameters used in physical simulations.}}
    \label{tab:parameters}
\end{table}

\begin{figure}[!htb]
\centering
\includegraphics[width=1.0\textwidth]{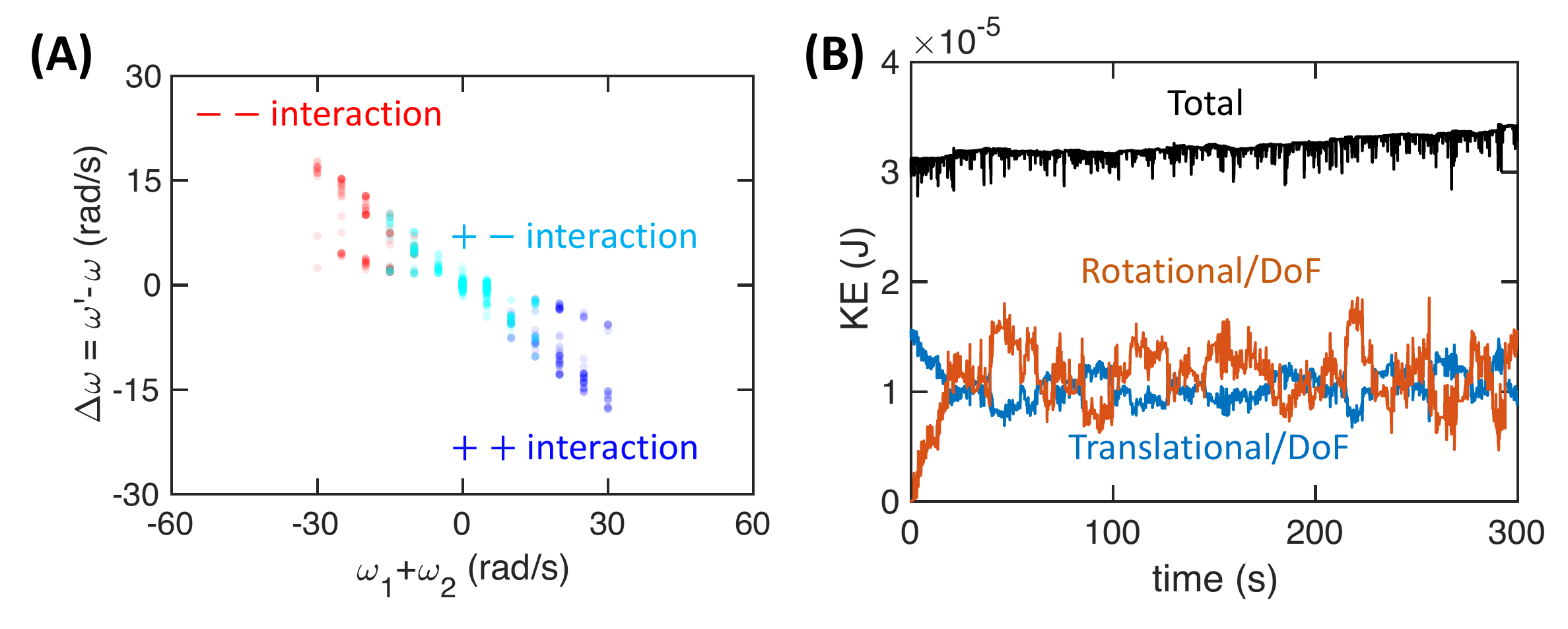}
\caption{\textbf{Simulation test.} (A) Change of angular velocity from simulation shows a negative proportionality with a slope $-0.5$, slightly larger than the $-0.38$ in experiment. (B) Without dissipation or injection of energy, $18$ spinners started with pure translational energy converts part of the translational energy into rotational energy and shows equipartition. The value shown in this panel shows the energy per spinner per degree of freedom. The slight increase of total energy is from accumulative numerical error, which can improve with finer time steps. See \href{https://www.dropbox.com/s/w4yb21ikjhta5c8/SI5_ffmpeg.mp4?dl=0}{SI5.mp4} for video.}
\label{fig:sim_test}
\end{figure}

For the spinner-spinner collision force, Each spinner is modeled as line segments connected to each other and uses the exact geometry of the spinner. The spinner-spinner interaction is evaluated as the sum of all pairwise interactions between the line segments of the two spinners. The line-line interaction uses spring-dash model, which regards the strain as the virtual overlap of the two line segments $F_{coll}=k_s \delta +k_d \delta v_n$ where $v_n$ is the relative velocity projected in the normal direction (Fig.\ref{fig:simSetup}C). $k_s$ and $k_d$ are phenomenological parameters such that the coefficient of restitution and the collision pattern $\Delta \omega = -\beta (\omega_1+\omega_2)$ matches with the experiment (Fig.\ref{fig:sim_test}A). In simulations, we use $k_s=10^4$ N/m,$ k_d=2\times 10^5$ N s/m$^2$.

\subsubsection{B. Numerical method}
The code was first developed in MATLAB for visual convenience and then manually compiled into C++ for efficiency. The numerical scheme uses an sympletic integrator, velocity-Verlet \cite{swope1982computer} to reduce accumulated numerical errors. The time step uses $10^{-4}$ second considering the largest Jacobian related to the collision is $\sim 10^4$ second$^{-1}$ as we choose the collision elasticity to be $10^4$ N/m. The elasticity is phenomenonlogical and yet physically realistic that the collision result is insensitive to the elasticity value given the order of magnitude is $\sim 10^4$ N/m.

As a test, we evaluate a simulation at equilibrium condition (without air current force, rotational or translation drag or drive) where $18$ non-active spinners started with pure translational motion. Over time, a portion of translational energy gradually converts to rotational energy and eventually shows equipartition in rotational (1 degree of freedom) and translational energy (2 degrees of freedom), i.e. $\frac{1}{2}m\langle v^2_i\rangle_i /2= \frac{1}{2} I \langle\omega_i^2\rangle_i/1$ \cite{nichol2012equipartition}. It is an interesting feature that with concave geometry, there can be tangential interaction without having dissipative forces \cite{liu2020oscillating}.

\subsubsection{C. Results}
We first simulate the collective behavior of the $18$-spinner and $36$-spinner systems. All simulation runs for 500 s to be consistent with the experiments. See \href{https://www.dropbox.com/s/w4yb21ikjhta5c8/SI5_ffmpeg.mp4?dl=0}{SI5.mp4} for the simulation. For each spin ratio, an ensemble of $10$ simulations is done to evaluate the kinetic energies' and mixing time's dependence on spin ratio. Without fine tuning, the results match with experiments except for small quantitative deviation. Further, the small vortices for regimes without global circulation also show motile vortices as we observe in experiments (Fig.\ref{figS09}). We further use simulations explore cases with $N=24,30$ to search for the boundary between the regimes with and without global circulation (Fig.3E,F in the main text).

\begin{figure}[!htb]
\centering
\includegraphics[width=0.8\textwidth]{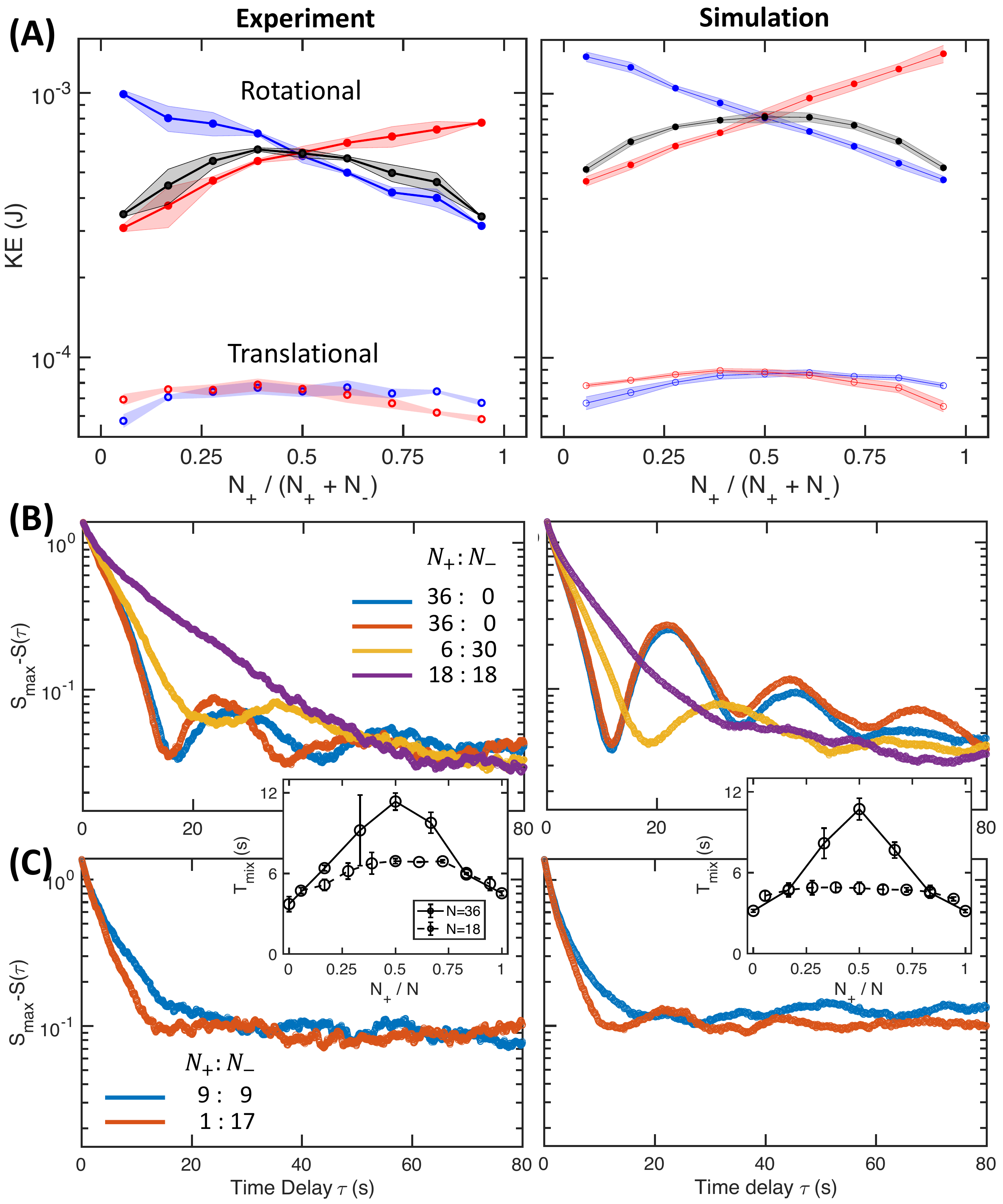}
\caption{\textbf{Simulation result.} Comparison of experiment and simulation in terms of (A) kinetic energies' dependence on spin ratio, (B) the mixing dynamics of a $36-$spinner system, and (C) the mixing dynamics of a $18-$spinner system.}
\label{fig:simResult}
\end{figure}

\clearpage
\subsection{VIII. Supplementary Movies}

\href{https://www.dropbox.com/s/qbyxp9zg0iblhmq/SI1_ffmpeg.mp4?dl=0}{\textbf{[SI1] Introduction to the spinners}}

This movie first introduces an individual spinner part by part, then shows the individual motion and collective motion from a perspective view, and finally shows two typical pairwise interactions, being the same-spin, and opposite-spin interactions.\\

\href{https://www.dropbox.com/s/upmd9s47xk7fssc/SI2_ffmpeg.mp4?dl=0}{\textbf{[SI2] Collective motion of the spinners}}

This movie shows the collective motion with different spin-ratios ($N_+:N_- = 0:1$, $0.5:0.5$, and $1:0$) at different number densities ($N=N_+ +N_- = 18$ and $36$). The motion is displayed at both real time and three times faster.\\

\href{https://www.dropbox.com/s/t96o32yb8btcos6/SI3_ffmpeg.mp4?dl=0}{\textbf{[SI3] Mixing of marked spinners}}

This movie first visually demonstrates the method we use to obverse the mixing process, which leads to the evaluation of entropy change shown in the manuscript. The movie then shows the collective motion with different spin-ratios ($N_+:N_- = 0:1$, $0.5:0.5$, and $1:0$) at different number densities ($N=N_+ +N_- = 18$ and $36$). The spinners are also marked as in the method demonstration and the videos are displayed at both real time and three times faster.\\

\href{https://www.dropbox.com/s/i3opuwfvxk2j154/SI4_ffmpeg.mp4?dl=0}{\textbf{[SI4] Fractal spinner}}

This movie shows the motion of two typical configurations composed of four spinners (two up spinners and two down spinners) in a confining ring (Fig. 4 in the main text).\\

\href{https://www.dropbox.com/s/w4yb21ikjhta5c8/SI5_ffmpeg.mp4?dl=0}{\textbf{[SI5] Simulation}}

This movie first shows the closeup of simulation at a speed 10 times slower than the real time. The movie then shows the collective motion reproducing the experiments shown in SI2. Finally, the movie shows the equipartition of energy holds in a simulation where all energy injection and dissipation are turned off.


\end{document}